\documentclass{aa}  
\usepackage{natbib,twoopt}
\usepackage[breaklinks=true]{hyperref}
\usepackage{subfig}
\hypersetup{
    colorlinks=true,
    linkcolor=blue,
    citecolor=blue,
    filecolor=blue,
    urlcolor=blue
}

\usepackage{graphicx}
\usepackage{txfonts}
\begin{document}

   \title{RedDots: Multiplanet system around M dwarf \\ GJ 887 in the solar neighborhood}


   \author{C.~Hartogh\inst{\ref{IAG}, \ref{STA}}  
          \and S.~V.~Jeffers\inst{\ref{TLS}}
          \and S.~Dreizler\inst{\ref{IAG}}          
          \and J.~R.~Barnes\inst{\ref{OUM}}
          \and C.~A.~Haswell\inst{\ref{OUM}}
          \and F.~Liebing\inst{\ref{PA}}
          \and A.~Collier~Cameron\inst{\ref{STA}}
          \and P.~Gorrini\inst{\ref{IAG}}          
          \and F.~Del~Sordo\inst{\ref{ICE},\ref{INAF},\ref{SNS}}          
          \and P.~Cort\'es-Zuleta\inst{\ref{STA}}          
          }

    \institute{Institut f\"ur Astrophysik und Geophysik (IAG), Universit\"at G\"ottingen,
            Friedrich-Hund-Platz 1, 37077 G\"ottingen, Germany
            \label{IAG}
            \and
            Centre for Exoplanet Science, SUPA, School of Physics and Astronomy, University of St Andrews, North Haugh, St Andrews KY16 9SS, UK, \email{ch395@st-andrews.ac.uk}            
            \label{STA}
            \and
            Th\"uringer Landessternwarte Tautenburg, Sternwarte 5, D-07778 Tautenburg, Germany 
            \label{TLS}
            \and
            School of Physical Sciences, The Open University, Walton Hall, Milton Keynes, MK7 6AA, UK  
            \label{OUM}
            \and
            Private Astronomer 
            \label{PA}
            \and
            Institut de Ci\`encies de I'Espai (ICE-CSIC), Campus UAB, Carrer de Can Magrans s/n, 08193 Cerdanyola del Vallès, Barcelona, Catalonia, Spain 
            \label{ICE}
            \and
            INAF, Osservatorio Astrofisico di Catania, via Santa Sofia, 78 Catania, Italy
            \label{INAF}
            \and
            Scuola Normale Superiore, Piazza dei Cavalieri, 7 56126 Pisa, Italy
            \label{SNS}           
            }

   \date{Received...}

 
  \abstract
   {GJ 887 is a bright M dwarf in the solar neighborhood with two currently reported nontransiting exoplanets with periods of $9~\mathrm{d}$ and $21~\mathrm{d,}$ along with an additional unconfirmed signal at $50~\mathrm{d}$.
   We reanalyzed the system with 101 new HARPS and 12 new ESPRESSO radial velocities (RVs) secured with a cadence to confirm or refute the origin of the $50~\mathrm{d}$ signal. To do so, we searched for signals related to stellar activity in photometric data and spectroscopic indicators.
   We modeled the stellar activity in the RVs with Gaussian processes (GPs).
   With the Bayesian analysis, we confirmed a four-planet model,   including the two previously known planets at periods of $9.2619\pm0.0005~\mathrm{d}$ and $21.784\pm0.004~\mathrm{d,}$ as well as two newly confirmed exoplanets: an Earth-mass planet, with a $4.42490\pm0.00014~\mathrm{d}$ period and a sub-meter-per-second amplitude, and a super-Earth with a $50.77\pm0.05~\mathrm{d}$ period located in the habitable zone (HZ).
   This super-Earth is the second closest planet in the HZ, after Proxima Cen b. We found an additional signal in a 2:1 resonance with the $4.4~\mathrm{d}$ planet at $2.21661\pm0.00010~\mathrm{d}$ with an amplitude of $0.37\pm0.09~\mathrm{m/s}$, which could be related to an additional planet. However, other explanations of its origin are also plausible. This signal remains a candidate, as further investigation is required to confirm its true nature. If the signal is caused by a planet, its minimum mass would be half that of Earth. 
   We measured the stellar rotation period with the characteristic periodic timescale of the GP. We found a period of $38.7\pm0.5~\mathrm{d}$, which is consistent with the rotation period determined from photometry and other activity indices.
   }

   \keywords{techniques: radial velocities --
                stars: activity --
                stars: individual: GJ 887 }

   \maketitle
%

\section{Introduction}
Radial velocity (RV) surveys with stabilized high-precision spectrographs, 
such as High Accuracy Radial velocity Planet Searcher (HARPS; \citealt{HARPS}), Echelle SPectrograph for Rocky Exoplanets and Stable Spectroscopic Observations (ESPRESSO; \citealt{ESPRESSO}), and Calar Alto high-Resolution search for M dwarfs with Exoearths with Near-infrared and optical Echelle Spectrographs (CARMENES; \citealt{CARMENES}), together with space-based photometric surveys such as  Kepler (\citealt{Kepler}) and Transiting Exoplanet Survey Satellite (TESS; \citealt{TESS}) have played key roles in the detection and characterization of more than 5,000 confirmed exoplanets. The best spectrographs produce RV uncertainties $ < 1~\mathrm{m\,s^{-1}}$, leading to a steady flow of low-mass planet detections. Low-mass planets in the habitable zone (HZ) of their host stars 
are particularly exciting and M dwarfs are advantageous for finding such planets since their reflex RV amplitudes are higher and the HZ is located at shorter orbital periods than for Sun-like stars.  Across all stellar types, there are 70 planets in the HZ according to the Habitable Worlds Catalog\footnote{\url{https://phl.upr.edu/hwc}}.

One of the main limitations in detecting these planets is the magnetic activity of their host stars. Spots and plage are regions in the stellar atmosphere caused by magnetic activity. They create flux inhomogeneities on the visible stellar surface that result in asymmetries that cause apparent Doppler shifts in the location of the spectral line center. Consequently, on timescales of a few rotations of the star, spots and plages can be misinterpreted as Doppler reflex motions of the star caused by exoplanets, while stellar variability appears as correlated noise that blurs exoplanetary signals. Since spots and plage have finite lifetimes, both the amplitudes and phases of their signals in the RVs are incoherent. The amplitudes of the apparent Doppler shifts of spots and plage are usually on the order of $1-10~\mathrm{m/s}$ \citep{Crass2021}. M dwarfs show a wide range in  magnetic activity ranging from inactive to very active \citep[e.g.,][]{Jeffers2018,Jeffers2022}. On multiple occasions, the reported planet signals were questioned in follow-up papers that analyzed the same systems with additional data and different activity models. The planetary status of reported planet signals was downgraded in several cases in follow-up publications (e.g., GJ 667 C d \citealt{Robertson2014}, GJ 832 c \citealt{Gorrini2022}) because the signals were attributed to stellar activity. We modeled the activity from correlated noise models with a Gaussian process (GP) regression (e.g., \cite{Haywood2014}). Other stellar phenomena that affect RV measurements include p-mode oscillations and granulation, but these are less prominent in M dwarfs than in FGK-stars and they can be reduced significantly by nightly binning of multiple observations \citep{Dumusque2011}.

\citet{Jeffers2020} reported two planets on orbits of $9~\mathrm{d}$ and $21~\mathrm{d}$ around the M1 dwarf GJ 887. 
An additional signal at $50~\mathrm{d}$ was identified but it was unclear whether this was caused by stellar activity or a planet.  
Using a GP model they found the likelihood
of the two-planet and three-planet model was statistically indistinguishable.
\citet{Jeffers2020} found GJ\,887 was less magnetically active than other stars of similar spectral type. The combination of the quiet, nearby star, confirmed planets close to the inner edge of the HZ, and the possibility of a third planet within the HZ means 
GJ\,887 is particularly interesting for further characterization. If the $50~\mathrm{d}$ signal is due to a planet, the system would be a prime candidate for atmospheric characterization, with such  proposed imaging missions such as the Habitable Worlds Observatory \citep[HWO;][]{Mamajek2024}
or interferometry missions such as Large Interferometer For Exoplanets \citep[LIFE;][]{LIFE} 
due to its brightness and proximity to the sun.

We present new RV data on  GJ 887, secured by the RedDots collaboration, which we analyze along with archival data.
RedDots uses a nightly observing cadence strategy which has led to detections around our nearest stellar neighbor Proxima Centauri \citep{Anglada-Escude2016} and various other nearby stars, such as GJ 1061 \citep{Dreizler2020} and GJ 887 \citep{Jeffers2020}.  In Section \ref{Stellar Properties}, we present the properties of the star. In Section \ref{Observational Data}, we describe observations. The methods to analyze the observations are discussed in Section \ref{Methods} and the results are presented in Section \ref{Results}. In Sections \ref{Discussion} and \ref{Summary}, we discuss and summarize the results. 

\section{Stellar properties}\label{Stellar Properties}

GJ 887, also known as HD 217987 and Lacaille 9352, has an effective temperature of $3688\pm86~\mathrm{K}$, corresponding to an M1V spectral type \citep{Mann2015}. It has solar-like metallicity, [Fe/H]=$-0.06\pm0.08$, and with distance of $d=3.2877~\mathrm{pc}$ and $V=7.39$, it is one of the brightest M dwarfs in the sky. Table \ref{tab:stellarparams} summarizes the stellar parameters. In the interval 1998 to 2018, GJ\,887 exhibited low starspot coverage, low photometric variability, low activity in the Ca II H and K lines, and low H$_\alpha$ activity \citep{Jeffers2020}. \citet{Jeffers2020} found periodicities in activity indicators around $38~\mathrm{d}$ and $55~\mathrm{d}$, however, due to the low activity level the rotation period ($P_{rot}$), could not be determined.

\begin{table}
    
    \centering
    \caption{Stellar parameters of GJ 887}
    \begin{tabular}{ll}
        \hline
        \textbf{Parameter} & \textbf{Value}  \\
        \hline\hline
        Spectral Type\tablefootmark{a}                    & M1V                  \\
        Parallax [mas]\tablefootmark{b}                  & $304.135\pm0.020$            \\
        Distance [pc]\tablefootmark{b}                   & $3.2877$           \\
        V Magnitude\tablefootmark{b}                       & $7.39$                \\
        Mass [$\mathrm{M}_\odot$]\tablefootmark{a}       & $0.495\pm0.049$                 \\
        Metallicity [Fe/H]\tablefootmark{a}               & $-0.06\pm0.08$               \\
        $T_{\text{eff}}$[K]\tablefootmark{a}              & $3688\pm86$              \\
        Luminosity [$\mathrm{L}_\odot$]\tablefootmark{a}  & $0.0368\pm0.0040$            \\
        Radius [$\mathrm{R}_\odot$]\tablefootmark{a}      & $0.468\pm0.022$             \\
        $v \sin i$ [km/s]\tablefootmark{c}               & $<2.5$                      \\
        Stellar Age [Gyr]\tablefootmark{a}              & $2.9_{-2.2}^{+8.0}$                     \\
        $\log g$\tablefootmark{b}                        & $4.156_{-0.038}^{+0.040}$          \\
        \hline
    \end{tabular}
    \tablebib{\tablefoottext{a}{\cite{Mann2015}}\tablefoottext{b}{\cite{GAIA}}\tablefoottext{c}{\cite{Browning2010}}}

    \label{tab:stellarparams}
\end{table}

\section{Observational data}\label{Observational Data}

\subsection{Spectroscopy}

We used a 
total of 
850 spectra from HARPS and 19 from the ESPRESSO. 
There are other publicly available data:
75 from Keck, 38 from UCLES and 38 from PFS, but we elected not to use them.  
UCLES (median uncertainty: $1.87~\mathrm{m/s}$), Keck (median uncertainty: $2.15~\mathrm{m/s}$), and PFS (median uncertainty: $1.02~\mathrm{m/s}$) all show a significantly lower precision than the HARPS and ESPRESSO data.

\subsubsection{HARPS}
HARPS is a fiber-fed echelle spectrograph installed at a 3.6m telescope located at La Silla observatory in Chile. It reaches a spectral resolution of 115,000 on 72 echelle orders wavelengths between 3800~$\AA$ and 6900~$\AA$.
The observations are listed 
in Table~\ref{tab:RVproposals} and the RV time series is given in Table~\ref{tab:RVdata}. Approximately half were obtained in high-cadence sequences spread over six nights: 402 by 0101.D-0494(B) + 90 by 191.C-0505(A). This sampling is unsuitable for the periods we wish to examine. There are 
176 archival spectra and 182 RedDots spectra with suitable sampling. After nightly binning of the data, 277 RVs remain, 
101 of which were obtained after the previous study of \cite{Jeffers2020}. 
The spectra used by \cite{Jeffers2020} are indicated 
in Table \ref{tab:RVproposals}.

The RVs were computed by template-matching with \texttt{serval} \citep{serval}, which fits the RVs order-by-order. For M dwarfs template matching, by fitting the full observed spectrum, better accounts for line blending and achieves higher RV precision compared to the cross-correlation-function (CCF) method \citep{Anglada-Escude2012}, which benefits from isolated lines that are more common in FGK-stars \citep{Anglada-Escude2012}.
As a first step, \texttt{serval} builds a high signal-to-noise co-added template from all 850 HARPS spectra and uses this to compute the order-by-order RV, which is then averaged to the total RV.
Additionally, the following activity indicators were computed by \texttt{serval}: line indices (Na D$_1$, Na D$_2$, and H$_\alpha$),
differential line width (dLW), which measures variations of the average line profile and the chromatic index (CRX), which is the wavelength dependence of the RVs measured by fitting a line to the individual order velocities.  
We also used the CCF full width half maximum (FWHM) and CCF bisector inverse span (BIS) \citep{Queloz2001} from the ESO-DRS pipeline. All of these indicators are helpful for determining $P_{rot}$ and to distinguish between planet reflex motions and stellar activity artifacts in the RVs.

The optical fiber change in 2015 and the warm-up of the instrument during the covid pandemic in 2020 both caused instrumental offsets in the RVs and changes in the instrumental jitter properties. Accordingly, 
we  separated the HARPS data into three different blocks with: 88 pre fiber change RVs (HARPSpre, median uncertainty: $0.44~\mathrm{m/s}$), 164 post fiber change RVs (HARPSpost, median uncertainty: $0.87~\mathrm{m/s}$), and 25 post warm-up RVs (HARPSpw, median uncertainty: $0.61~\mathrm{m/s}$).

\subsubsection{ESPRESSO}
ESPRESSO \citep{ESPRESSO} is a fiber-fed, cross-dispersed echelle spectrograph located in the Combined Coudé Laboratory at Paranal Observatory in Chile. ESPRESSO, installed on the 8.2-m Unit Telescopes (UTs) of the Very Large Telescope (VLT), provides a resolving power of $R \sim 138{,}000$ over the wavelength range 3780–7890~$\AA$. The ESPRESSO spectra display excellent precision (median uncertainty: $0.17~\mathrm{m/s}$), but after the nightly binning, only 12 spectra remained that were spread over a span of three years. We used RVs and uncertainties from the standard CCF procedure of the ESO-DRS pipeline.

\subsection{Photometric data}

To search for transiting planets we used Sectors 2, 28, and 69 of the TESS mission \citep{TESS}. Each sector covers $27~\mathrm{d}$ of continuous observations with a short break in the middle of the sector. GJ\,887 is 
magnitude 
5.56 in the TESS band, which leads to a median uncertainty of 0.00011 on the normalized flux of the pre-search data conditioning simple aperture photometry (PDCSAP) light curves. This is sufficient for detections of transiting planets with Earth radius, which would have a transit depth of 0.00038 around GJ 887. For a photometric determination of the stellar rotation period, we employed data from the All-Sky Automated Survey (ASAS; \cite{ASAS}). We used 523 observations by ASAS spanning from 2001 to 2009 with a median precision of 0.03 on the normalized flux. The light curve is shown in Figure \ref{fig:ASASTS}.

\section{Methods}\label{Methods}

\subsection{Periodogram methods}
Periodograms are a valuable tool for identifying periodic signals in RV data. As a first step, we used the generalized Lomb Scargle \citep[GLS,][]{Zechmeister2009} periodograms as a blind search for sinusoidial signals in the RVs. Additionally, we used stacked Bayesian generalized Lomb Scargle  \citep[sBGLS,][]{Mortier2017} periodograms to identify coherent signals (stable over the observing time) induced by planet reflex motions of the star and incoherent signals induced by stellar activity artifacts in the RVs. Furthermore, we used $\ell_1$-periodograms \citep{Hara2017}, which search for multiple Keplerian signals simultaneously, naturally down-weighting signals caused by noise or aliasing as the inclusion of an additional signal is penalized. Each of the three methods was applied to the RVs. Applying the GLS to the activity indicators offers a good first insight into the origin of signals in the RV data.

\subsection{Modeling}\label{Modelling}

For modeling and characterizing exoplanets in the presence of stellar activity, a sophisticated analysis scheme is required. 
We used the \texttt{juliet} package \citep{Espinoza2019}, in which a variety of routines for exoplanet characterization are implemented. The Keplerian RV models are implemented with the \texttt{RadVel} \citep{Fulton2018} package. For modeling the stellar activity, we used GP. Therefore, we employed the quasiperiodic kernel from \texttt{george} \citep{Ambikasaran2015} with a covariance function, 
\begin{equation}
    k(\tau)=\sigma_i^2\exp\left(-\alpha\tau^2-\Gamma\sin^2\left[\frac{\pi\tau}{P_{rot}}\right]\right),
\end{equation}
with the amplitude of the signal for each specific HARPS interval, $\sigma_i$, the inverse length scale, $\alpha$, and the amplitude of the oscillatory part, $\Gamma$. The mean lifetime of the spots $l$ can be estimated from $l=1/\sqrt{\alpha}$. We compare the activity model of the QP kernel to other activity models that we ran with the Simple Harmonic Oscillator (SHO) and double-SHO (dSHO) kernels \citep[e.g.,][]{Kossakowski2021} from the \texttt{celerite} package \citep{Foreman-Mackey2017}. For the posterior estimation, we used the dynamic nested sampling scheme \texttt{dynesty} \citep{dynesty}. This is different from the 
Markov chain Monte Carlo (MCMC) technique used in the previous study \citep{Jeffers2020} since \texttt{dynesty} yields the Bayesian evidence, $\mathcal{Z}$, directly and is less likely to be trapped in local likelihood maxima. 
There is strong (moderate) evidence for one model in preference to another if $\Delta\log\mathcal{Z}>5 \,\,( \Delta\log\mathcal{Z} > 2.5)$
\citep{Trotta2008}. This enables us to quantitatively compare results with different numbers of planets, as well as for circular and eccentric orbits and combinations thereof in multi-planet systems. For each RV interval, we fit a separate jitter and offset as well as a separate GP amplitude, $\sigma_i$. The  priors used are given in Table \ref{tab:priors}.

\subsection{Dynamic stability}
To determine the stability of the system, we used the Stability of Planetary Orbital Configurations Klassifier \citep[\texttt{SPOCK};][]{Tamayo2020}, a machine learning tool
to determine the probability of a configuration being stable. For posteriors of several planet configurations, we computed the probability of stability of each sample. We used these weights to determine weighted medians and confidence intervals for the reported parameter values.

\subsection{Transit search}
We searched for transits in the PDCSAP light curves of TESS. To account for stellar variability in the light curve, we clipped outliers over $5\sigma$ and flattened out the trend with an r-spline implemented in \texttt{W{\={o}}tan} \citep{Hippke2019b}. 
We applied the transit least-squares \citep[TLS;][]{Hippke2019a} technique on the flattened PDCSAP curves. The method is based on blindly fitting transit models for a given period grid. However, instead of uninformed box-shaped transit shapes \citep[BLS;][]{Kovacs2002}, a median transit model out of all known transit curves is used fitting the same parameters (period, phase, and depth) as BLS. For each period on the grid we can compute a signal detection efficiency (SDE) which is a measurement of the statistical significance of a planet transiting at the given period.

\section{Results}\label{Results}

\subsection{Estimation of the stellar rotation period}\label{sec:Prot}
In the first step, 
we searched for periodocities that appear in time series of photometric data and spectroscopic indices.

\subsubsection{Photometric rotation period}\label{sec:Photoperi}
\begin{figure}
    \centering
    \includegraphics[width=0.49\textwidth]{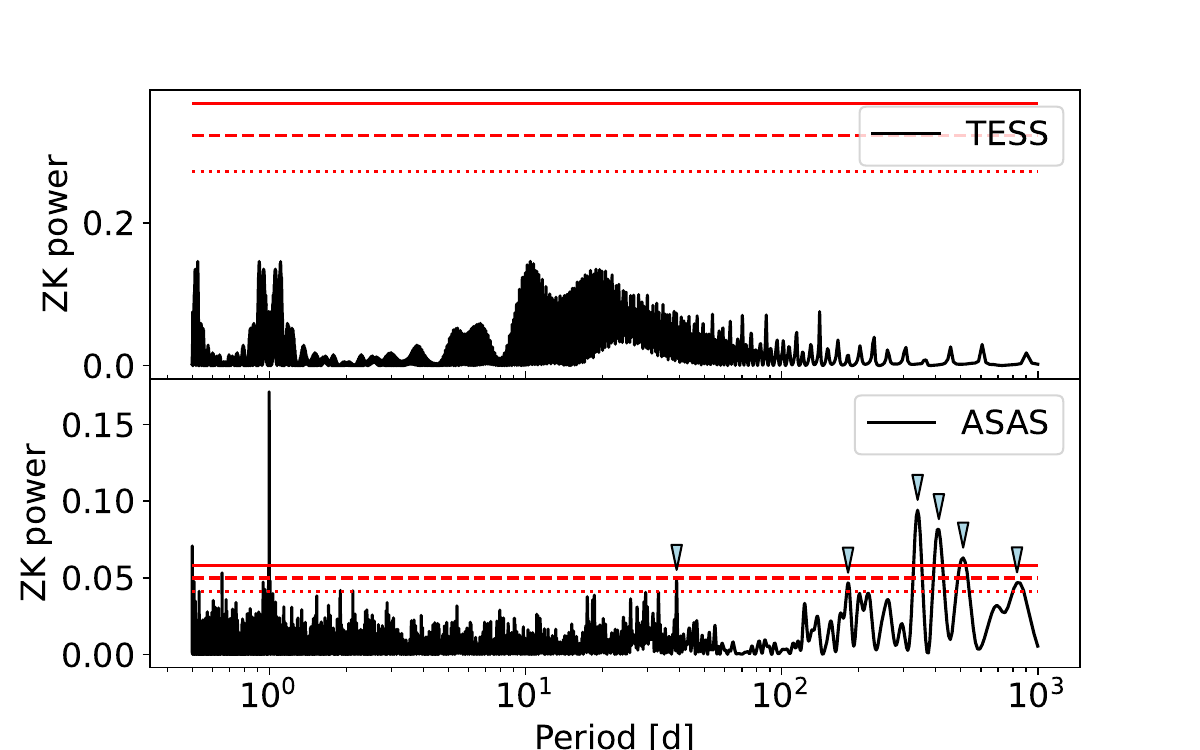}
    \caption{GLS periodograms of photometric TESS (upper panel) and ASAS (lower panel) data, including 0.1, 0.01, and 0.001 FAP levels in red. Significant peaks with periods higher than the 1~d sampling peak are highlighted in cyan triangles.}
    \label{fig:photoperi}
\end{figure}

We used 
data from ASAS and TESS to estimate the rotation period photometrically. After binning the (unflattened) TESS data to daily chunks, we computed the GLS periodogram for both datasets (Figure~\ref{fig:photoperi}). In the TESS periodogram, 
no signals are seen to reach the 0.1 FAP level. This is expected since the sampling of the data in three $27~\mathrm{d}$ long sectors, which makes the periodogram unreliable for longer periods of $>\sim12~\mathrm{d}$ \citep{Boyle2025}. The sampling of the TESS data is therefore suboptimal for detecting rotations of M stars that typically have longer rotation periods \citep{Suarez-Mascareno2016}.  In the more regularly sampled ASAS data (Fig. \ref{fig:ASASTS}), there is a prominent 1\,d sampling peak due to the window function, but multiple other signals show up in the periodogram. 
The periodogram shows several long period peaks at $340~\mathrm{d}$ and above which are likely caused by systematics or long-term signals of the star. The peak at $182~\mathrm{d}$ is likely a systematic at half a year.
The highest remaining peak is around $39~\mathrm{d}$. This could be related to stellar activity, and is a typical rotation period for an early M dwarf, while the periods exceeding $160~\mathrm{d}$ are not expected for M dwarfs  \citep{Suarez-Mascareno2016,Newton2018}.

\subsubsection{Spectroscopic rotation period}

To confirm the rotation period around $39~\mathrm{d}$ with spectroscopic indices, we computed their GLS periodograms.
In several indicators the long-term trend over 20 years is dominating the signal (e.g., Figure \ref{fig:longtermtrend}). Since our aim is to determine the rotation period, rather than the long-term trend, we need to account for it.

To do so, we subtracted the mean of all points within $600~\mathrm{d}$ of the observation from each data point.
This timescale was chosen to be significantly higher than the plausible rotation period
so that only long-term activity cycle trends would be filtered out.

\begin{figure}
    \centering
    \includegraphics[width=0.49\textwidth]{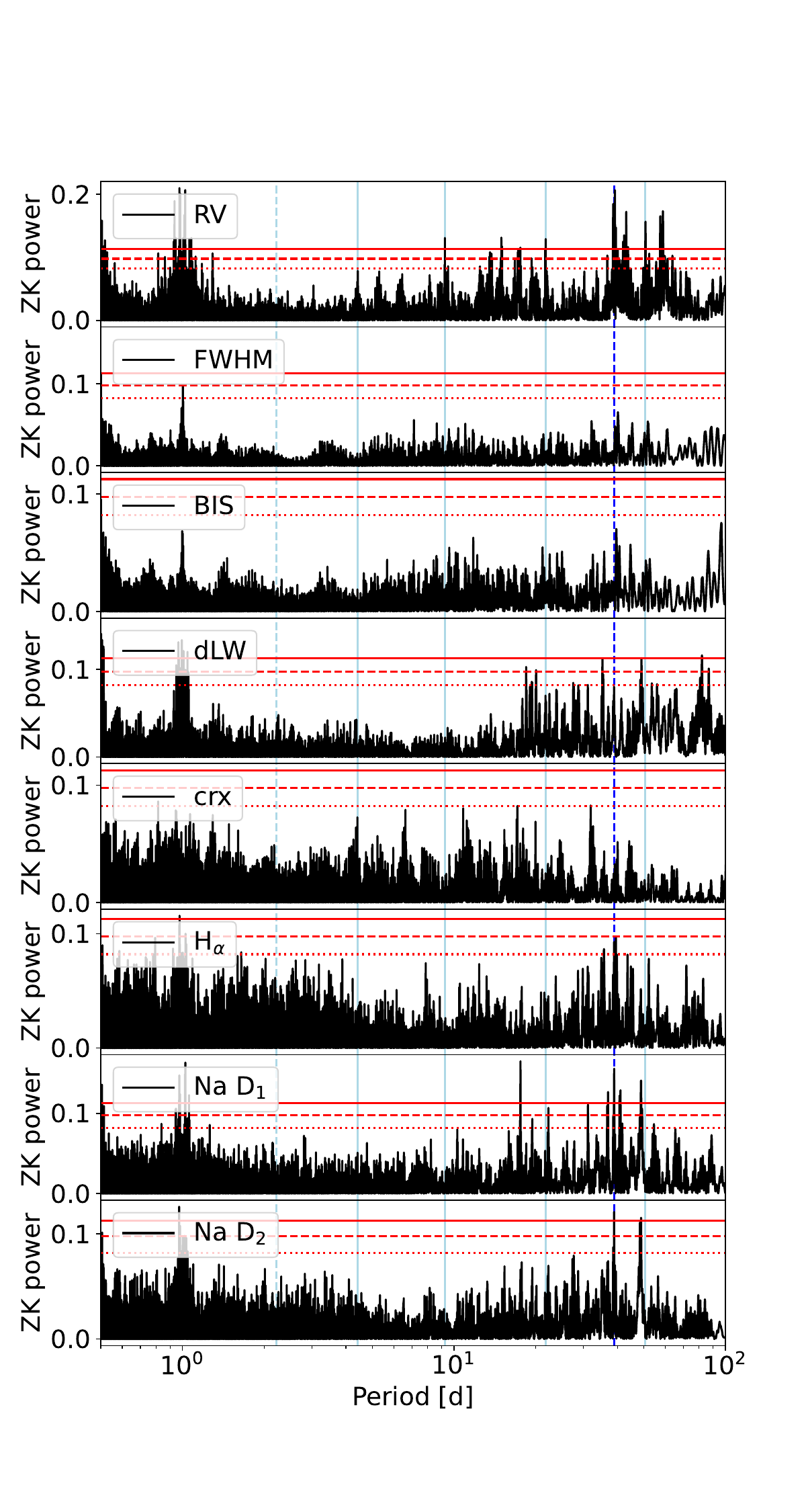}
    \caption{GLS periodograms of HARPS RV data and seven activity indicators described in the plots including 0.1, 0.01, and 0.001 FAP levels. The long-term trend was removed from all time series by subtracting the $600~\mathrm{d}$ average from each observation. The blue line indicates the rotation period determined by photometry. The cyan lines represent signals that we attribute to planets (solid) and candidates (dashed).}
    \label{fig:activityperi}
\end{figure}

Most of the activity indicators, RV, FWHM, BIS, dLW, $\mathrm{H}_\alpha$, Na D$_1$, and Na D$_2$, have a signal close to the 39\,d photometric period (Figure \ref{fig:activityperi}).
Although the FWHM and BIS $39~\mathrm{d}$ peaks do not reach the 0.1 FAP level, they still appear 
to be noteworthy, as the highest, apart from the 1\,d sampling artifact.
Some indicators show a peak around $19~\mathrm{d}$, which would be the first harmonic of the rotation period \citep[cf.][]{Barnes24}. The peaks at $49~\mathrm{d}$ in the dLW, Na D$_1$, and Na D$_2$ are unlikely to be aliases of the $39~\mathrm{d}$ signal, as their amplitudes do not match the alias pattern expected from the sampling \citep{Dawson2010}. Their consistently weaker power across most indicators argues against a rotational origin, but an activity-related explanation cannot be excluded. 

\begin{figure}
    \centering
    \includegraphics[width=0.49\textwidth]{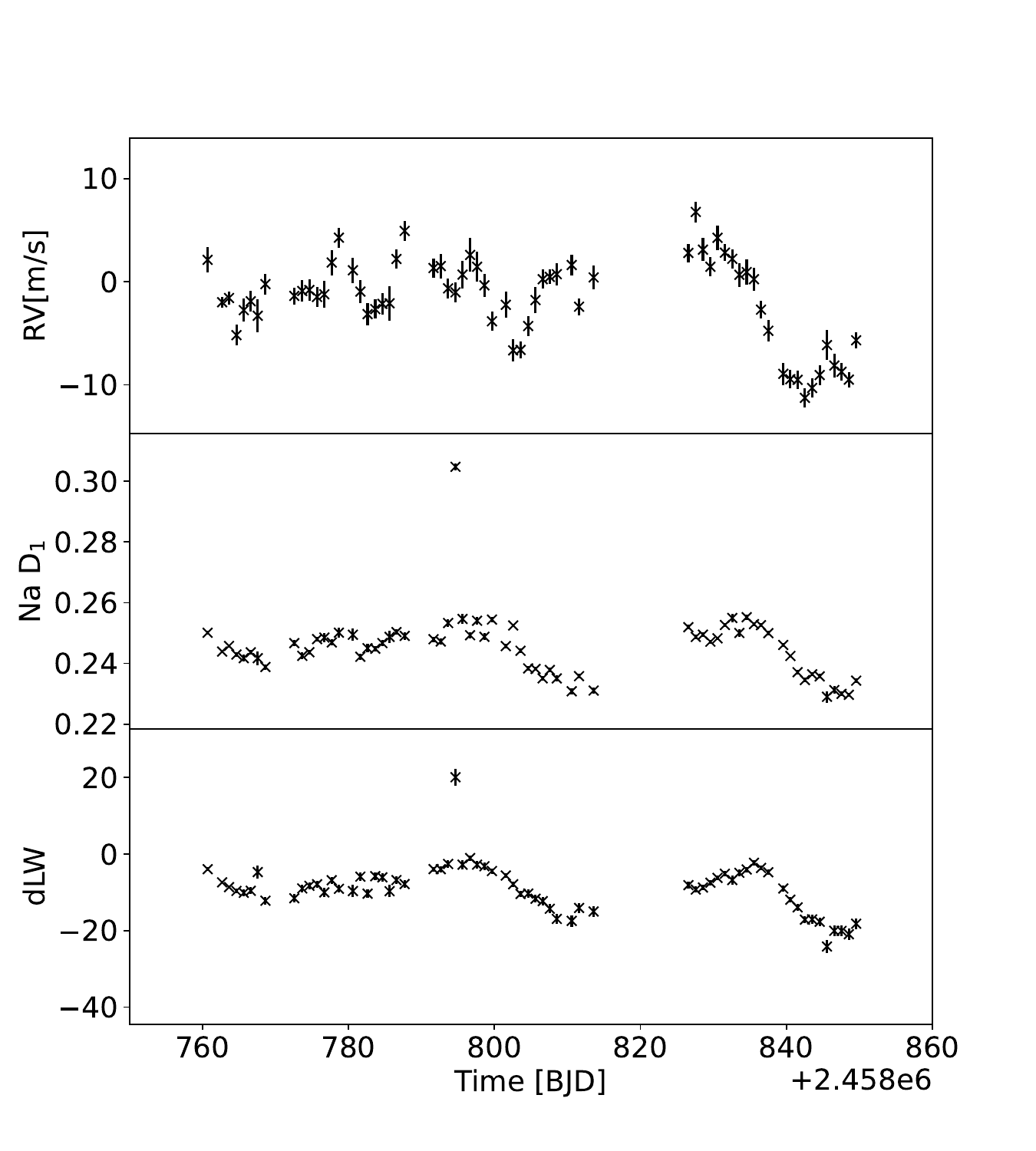}
    \caption{Time series of the active phase of GJ 887 in the RV and in Na D$_1$ and dLW indicators. The spectra were recorded during the 2019 RedDots observing campaign.}
    \label{fig:activephase}
\end{figure}

In the new RedDots observing campaign in 2021, the star showed strong activity in multiple activity indicators (e.g., differential line width, dLW, Na D$_1$) as well as in the RV itself (Fig. \ref{fig:activephase}). A long-term trend, attributable to an activity cycle, can be seen in those indicators (Fig. \ref{fig:longtermtrend}).

In one of the consecutively observed seasons, the star shows an activity outbreak (Figure \ref{fig:activephase}). The activity is most prominent in the RVs, Na D$_1$, and dLW. In these indicators, two clear minima and maxima separated by $39~\mathrm{d,}$ respectively, can be seen; these suggest increased stellar activity. 
This is additional evidence for a $39~\mathrm{d}$ rotation period with $19~\mathrm{d}$ as its harmonic. Nevertheless, in the GP models we investigate whether 
a rotation period around $39~\mathrm{d}$ is still valid when taking the planet signals into account.

\begin{figure}
    \centering
    \includegraphics[width=0.49\textwidth]{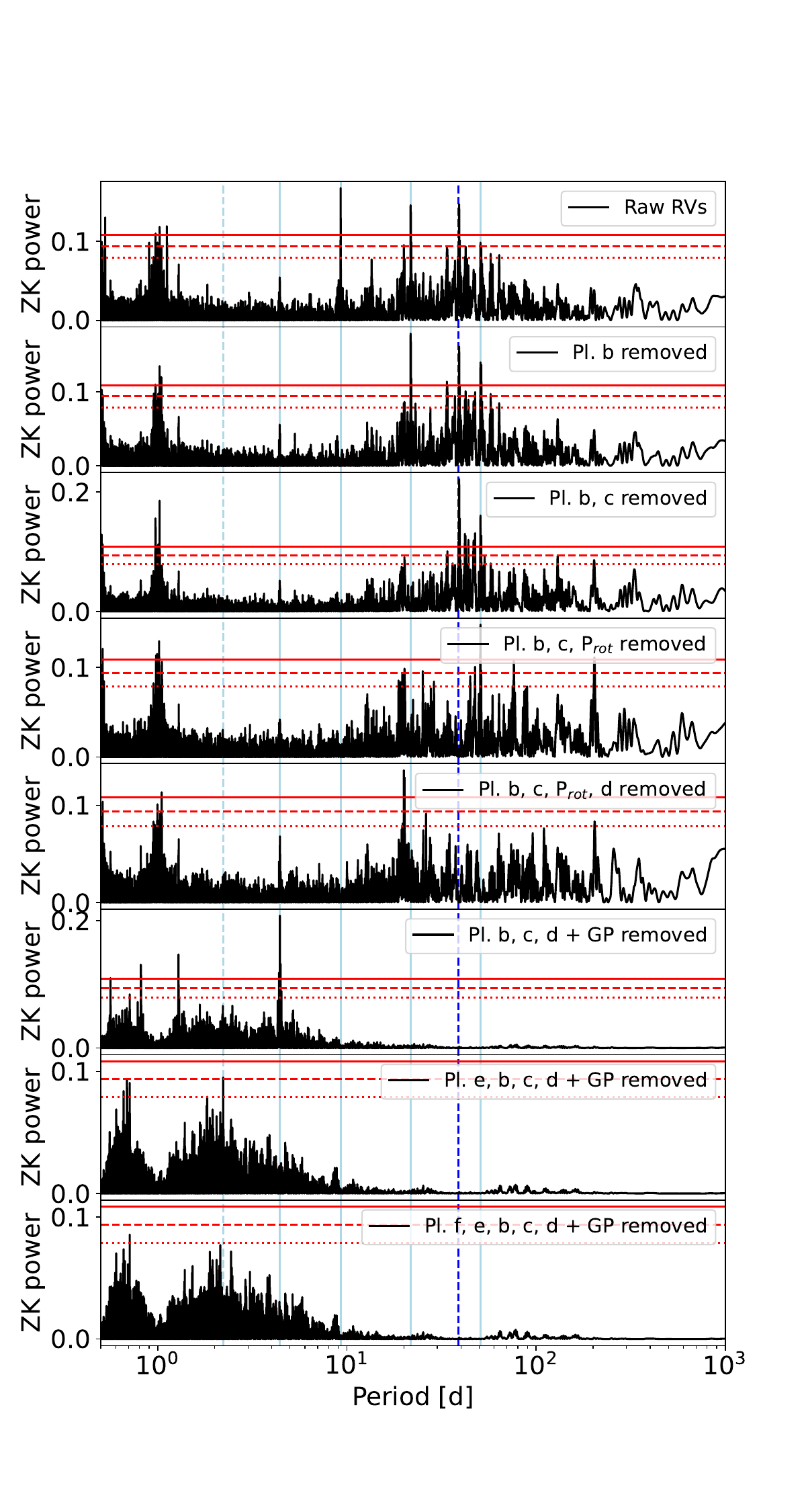}
    \caption{GLS periodogram of RVs 0.1, 0.01 and 0.001 FAP levels, shown in red. The blue line indicates the rotation period determined by photometry and spectroscopy. The cyan lines represent signals that we attribute to planets (solid) and candidates (dashed). The individual panels represent residuals of the respective model in the legend.}
    \label{fig:RVsignals}
\end{figure}

\subsection{RV signal search}\label{sec:RVsignalsearch}
In this section, we seek to determine the significance of signals in the RVs.
Figure~\ref{fig:RVsignals} shows 
the GLS periodogram of all HARPS and ESPRESSO RVs. The strongest peak belongs to planet b at $9.3~\mathrm{d}$ with a false alarm probability (FAP) of the signal of $5.8\times10^{-8}$. Thus, at this point, we removed the signal from the data. We modeled the new signal with a Keplerian in order to determine the component to subtract. Subtracting the $9.3~\mathrm{d}$ signal leaves the highest peak at $21.8~\mathrm{d}$ with a FAP of $8.0\times10^{-9}$, which belongs to planet c. Subtracting this signal leaves the $39~\mathrm{d}$ rotation period as the highest peak with an FAP of $3.9\times10^{-12}$. As we shown in the previous section, this belongs to the rotation period; however, for the time being, we chose to remove the signal with a Keplerian. This leaves a signal at $50.7~\mathrm{d}$ with an FAP of $1.6\times10^{-6}$ at the period of the planet candidate d. Subtracting this $50.7~\mathrm{d}$ signal leaves a peak at $19~\mathrm{d}$, which is half the rotation period and therefore unlikely to be caused by a planet. To avoid this effect we fitted the stellar activity with a quasiperiodic GP described in \ref{Modelling}. We fit the hyperparameters simultaneous with the planets and remove the predictive mean. The model with three planets and a GP reveals a signal at $4.4~\mathrm{d}$ with a FAP of $4.6\times10^{-11}$ in the residuals. Including the fourth Keplerian signal in the model leaves a signal at $2.2~\mathrm{d}$ with an FAP of $8.3\times10^{-3}$. Including this planet in the model as well leaves no significant signals. The highest remaining peak with FAP $3.6\times10^{-2}$ is at $1.4~\mathrm{d}$.

\begin{figure}
    \centering
    \includegraphics[width=0.49\textwidth]{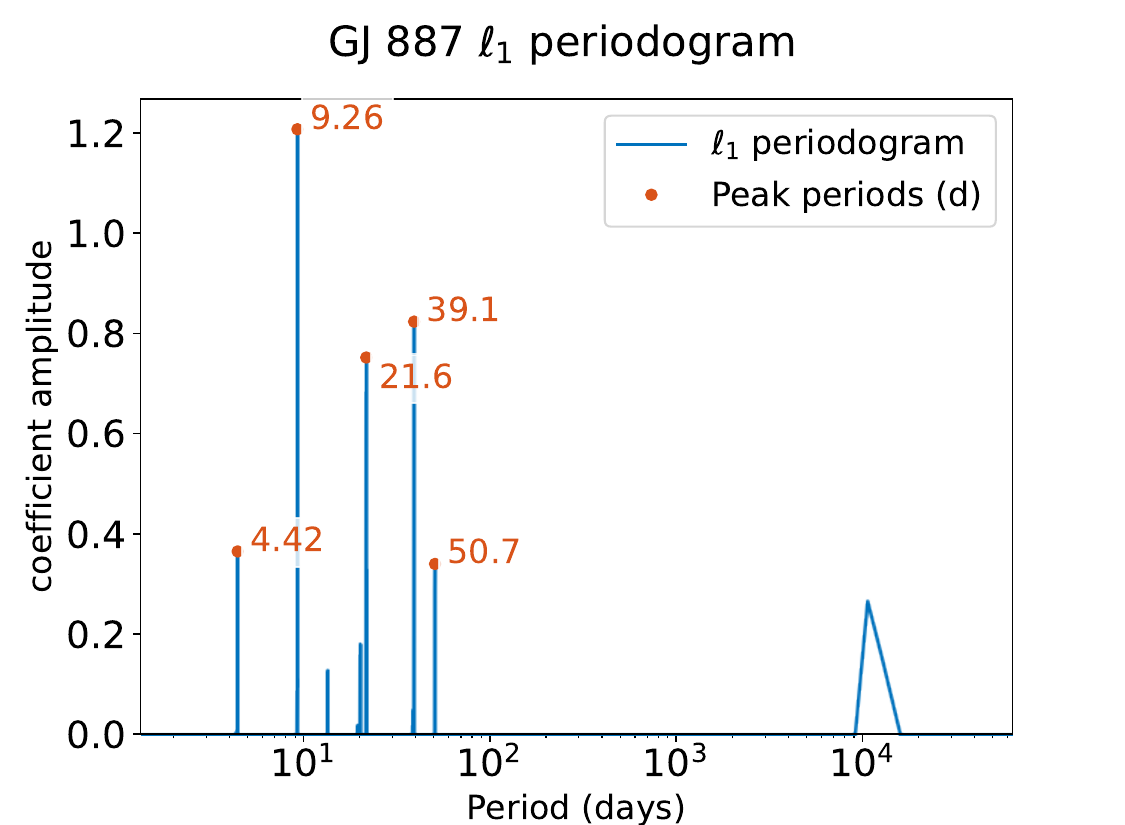}
    \caption{$\ell_1$-periodogram of all HARPS data. The model includes a white and a red noise model corresponding to the best likelihood as well as all signals exceeding a FAP of 0.1. Those signals are highlighted with red dots and their period.}
    \label{fig:l1peri}
\end{figure}

We also computed an $\ell_1$-periodogram (Fig. \ref{fig:l1peri}) for the HARPS RVs.
With this periodogram, we can model the noise with a combination of a white noise jitter term and a red noise Gaussian term with given correlation length. Then we divide the observations into a 60\% training and a 40\% test set following the method of \citet{Hara2020}. We generated a model, including all Keplerian signals with an FAP<0.1, and a noise term generated with given hyperparameters for white and red noise, on the training set. We then computed the likelihood of the data for the calculated model. For each set of hyperparameters we repeated this procedure 400 times and take the median of the likelihoods. Thus we assigned a likelihood to each set of hyperparameters. By exploring the parameter space for different hyperparameters and looking for the maximum likelihood, we obtained a good model for red and white noise.

We report the FAP values corresponding to the mean in $\log_{10}~\mathrm{FAP}$ of the best 20\% ranked models.
For these best parameter sets, 5 peaks in the $\ell_1$-periodogram remain significant, four of which belong to the periods of the planet candidates at $4.42~\mathrm{d}$ (FAP: $2.6\times10^{-4}$), $9.26~\mathrm{d}$ (FAP: $1.4\times10^{-10}$), $21.7~\mathrm{d}$ (FAP: $2.0\times10^{-10}$) and $50.7~\mathrm{d}$ (FAP: $5.2\times10^{-3}$); the fifth belongs to the rotation period at $39.2~\mathrm{d}$ (FAP: $1.1\times10^{-3}$).

\subsection{Modeling of RVs}\label{sec:modelRVs}
To determine the number of planets and estimate their orbital parameters, we ran Keplerian models.
In the data used in \cite{Jeffers2020}, the stellar contribution was minimal that models including and excluding a GP showed similar likelihoods. 
In the new RVs, however, GJ 887 shows strong activity. Since we have strong evidence for a rotation period around $39~\mathrm{d}$ we can include rotationally modulated GPs.
We tried the QP, SHO, and dSHO kernels for the GP models. The SHO failed to faithfully model the activity, yielding a worse $\log\mathcal{Z}$ and values in the characteristic oscillator frequency that correspond to periods of only a few days with uncertainty on the order of days. These values contradict the spectroscopic and photometric rotation periods and indicate overfitting, as the GP attempts to absorb short-term variability in the RVs. This shows that the SHO kernel does not provide an adequate description of the stellar activity. 
Models with the QP kernel showed slightly better $\log\mathcal{Z}$ compared to the dSHO kernel despite being less complex (four parameters rather than five). We adopt wide log-uniform priors on the GP period ($\mathcal{J}(10,100)$) to show that the estimated rotation period from photometry and spectroscopic indices agrees with the GP model. The other GP parameters are taken log-uniform distributions as well and can be found in Table \ref{tab:priors}.
The priors on the planetary periods were centered on the candidate signals reported by \citet{Jeffers2020} and those identified in our GLS and $\ell_1$ periodograms. This is common practice for Bayesian model comparison \citep{Gorrini2023,Dreizler2024,Basant2025} as shallow signals can be difficult to find for the algorithm in the presence of multiple planets and red and white noise in the RVs. We could make the argument that to assess the significance of a signal, a prior over the whole parameter space of the period (e.g., $\mathcal{J}(0.5,10000)$ d) must be considered. Therefore, we additionally report a corrected value for $\Delta\log\mathcal{Z}$, where the evidence of a model is multiplied by
\begin{equation}
    \frac{\log (P_1/P_2)}{\log (0.5~\mathrm{d}/10000~\mathrm{d})}
,\end{equation}
for the narrow period range of [$P_1$,$P_2$]. For the other planet parameters, we used uniform priors on the RV amplitude $\mathcal{U}(0.01,5)$ since signals of higher order of magnitude would be clearly visible by eye. The eccentricity is fixed to zero in circular models and has a prior of $\mathcal{U}(0,0.9)$ in eccentric models as higher eccentricities are very unlikely. The prior for $\omega$ is $\mathcal{U}(0,2\pi)$ and $t_0$ has a uniform prior with a width that matches the corresponding period. An extensive overview of all prior is given in Table \ref{tab:priors}.

\begin{table}[]
    \centering
    \caption{$\log\mathcal{Z}$ values of the results of all referenced GP models with the quasiperiodic (QP) kernel on binned data\tablefootmark{a}.}
    \begin{tabular}{lll}
        \hline Model &  $\log\mathcal{Z}$ QP &  $\Delta\log\mathcal{Z}$ QP \\
            \hline
            \hline  
    \textit{2c,4c,9e,21e,51c} (E) &   $-640.05\pm1.04$ &        +4.42   \\
       \textbf{\textit{4c,9e,21e,51c} (D)}  &   \textbf{$-644.47\pm0.84$} &        \textbf{0.00}  \\
        \textit{2c,4c,9e,21e} (F) &   $-649.47\pm2.64$ &      -5.00 \\
           \textit{4c,9e,21e} (C)&   $-650.95\pm0.38$ &       -6.48 \\
          \textit{9e,21e,51c} (B)&   -671.62 &      -27.15\\
              \textit{9e,21e} (A)&   -678.87 &      -34.4 \\
                  \textit{0P} &   -727.51 &      -83.04\\
                  \hline

    \end{tabular}
    \tablefoot{\tablefoottext{a}{Number describes orbital period, c and e indicate circular and eccentric orbits.}}
    \label{tab:resultsshort}
\end{table}

As a first step we tried to fit the two-planet model proposed \cite{Jeffers2020,Harada2025}. Both planets could be recovered clearly ($K_b=2.27_{-0.17}^{+0.17}~\mathrm{m/s}$ and $K_c=2.4_{-0.4}^{+0.4}~\mathrm{m/s}$ on orbits $P_b=9.2617_{-0.0005}^{+0.0005}~\mathrm{d}$ and $P_c=21.781_{-0.004}^{+0.005}~\mathrm{d}$) on eccentric orbits and the model is overwhelmingly preferred over a model without planets  with $\Delta\log\mathcal{Z}=48.64$ ($\Delta\log\mathcal{Z}=34.32$ when correcting for the choice of narrow priors). It is also moderately preferred over a model with both or one of the planets on a circular orbit. All Bayesian evidences are summarized in Table \ref{tab:results}. The key models are given capital letters in the text, and are listed in Table \ref{tab:resultsshort}. 

We compare the two-planet model (model A) with a three-planet model, including the $50~\mathrm{d}$ (model B) candidate and retrieve a model with a better evidence by $\Delta\log\mathcal{Z}=7.25$ ($\Delta\log\mathcal{Z}=0.34$ when correcting for the choice of narrow priors). For this planet, it is important to note that the GP affects and mitigates the signal. Comparing model A and model B for a non-GP model yields $\Delta\log\mathcal{Z}=19.20$ for the uncorrected and $\Delta\log\mathcal{Z}=12.29$ for the corrected case. Since the candidate was already identified  in \citet{Jeffers2020}, we named the candidate GJ 887 d. The planet in this case was fitted on a circular orbit, which led to a better $\mathrm{d}\log\mathcal{Z}$ than an eccentric orbit and its resulting RV semi amplitude was $K_d=1.8_{-0.4}^{+0.4}~\mathrm{m/s}$. In both models, the rotation period of the star was around $39~\mathrm{d}$. 

In the residuals of model B, a periodogram peak at $4.4~\mathrm{d}$ appears 
with a FAP of $7\times 10^{-12}$ (Fig.~\ref{fig:RVsignals} panel 6). This signal was detectable in the raw data (Fig. \ref{fig:RVsignals} panel 1 and Fig. \ref{fig:l1peri}); however, the peak in the GLS periodogram was less significant than peaks attributed to aliases or noise. Fitting a three-planet model with the candidate at $4.4~\mathrm{d}$ and planets from model A ($4.4~\mathrm{d}$ + A = model C) results in a significant improvement of $\Delta\log\mathcal{Z}=27.15$ ($\Delta\log\mathcal{Z}=18.22$ with corrected evidences) over A, which is strong evidence for the detection of a new planet. For this and all of the following models, the Bayesian evidences show values where variations of the $\Delta\log\mathcal{Z}$ values would qualitatively affect the interpretations on statistical significances. Therefore, we ran each model five times and reported the median $\log\mathcal{Z}$ and its uncertainty according to \citet{Nelson2020}. The medians and uncertainties can be found in Table \ref{tab:resultsshort}.
We call this new planet candidate GJ\,887\,e and it has an amplitude of $K_e=0.89_{-0.10}^{+0.09}~\mathrm{m\,s^{-1}}$.
We add the $50~\mathrm{d}$ candidate to model C (C + $50~\mathrm{d}$ = model D) and find an improvement of $\Delta\log\mathcal{Z}=6.48$ ($\Delta\log\mathcal{Z}=-0.43$) over model C, . The uncertainties of the evidences are both below 1. An overview of all posterior medians and confidence intervals of this model can be found in Table \ref{tab:results4p}.

\begin{table*}
\centering
    \caption{Table of posterior medians and $1\sigma$-intervals of model\,D\tablefootmark{a}. 
    }
\label{tab:results4p}
\begin{tabular}{lllll}

\hline Parameter &planet \textit{e} &planet \textit{b} &planet \textit{c} &planet \textit{d} \\
\hline\hline
$P$ [days] & $4.42490_{-0.00013}^{+0.00014}$ & $9.2619_{-0.0005}^{+0.0005}$ & $21.784_{-0.004}^{+0.004}$ & $50.77_{-0.05}^{+0.05}$ \\
$K$ [m/s] & $0.91_{-0.10}^{+0.10}$ & $2.00_{-0.17}^{+0.17}$ & $2.54_{-0.27}^{+0.27}$ & $1.7_{-0.4}^{+0.4}$ \\
$t_0$ [BJD] & $2453488.05_{-0.13}^{+0.13}$ & $2453492.75_{-0.27}^{+0.27}$ & $2453498.2_{-0.9}^{+0.9}$ & $2453496_{-4}^{+4}$ \\
$e$ & - & $0.14_{-0.06}^{+0.06}$ & $0.17_{-0.06}^{+0.06}$ & - \\
$\omega$ [rad] & - & $3.1_{-2.2}^{+2.2}$ & $3.5_{-2.4}^{+2.0}$ & - \\
$M \sin i [\mathrm{M}_{\oplus}]$ & $1.46_{-0.18}^{+0.19}$ & $3.9_{-0.5}^{+0.5}$ & $6.5_{-0.9}^{+1.0}$ & $6.1_{-1.4}^{+1.4}$ \\
$a$ [AU] & $0.0417_{-0.0015}^{+0.0014}$ & $0.0683_{-0.0024}^{+0.0022}$ & $0.121_{-0.005}^{+0.004}$ & $0.212_{-0.008}^{+0.007}$ \\
$T_{\mathrm{eq}}$ [K] & $544_{-20}^{+21}$ & $426_{-16}^{+16}$ & $320_{-12}^{+13}$ & $241_{-9}^{+10}$ \\
$S$ [$S_{\oplus}$] & $20.9_{-2.9}^{+4.0}$ & $7.8_{-1.1}^{+1.3}$ & $2.5_{-0.4}^{+0.4}$ & $0.81_{-0.12}^{+0.13}$ \\
$\theta$ [mas] & $12.7_{-0.5}^{+0.5}$ & $20.8_{-0.8}^{+0.7}$ & $36.7_{-1.3}^{+1.2}$ & $64.6_{-2.2}^{+2.1}$ \\
$C \times 10^{-9}$ & $103_{-7}^{+8}$ & $75_{-5}^{+6}$ & $63_{-4}^{+5}$ & $38.1_{-2.4}^{+2.8}$ \\
\hline
    \multicolumn{5}{c}{\textit{RV Parameters}} \\
    \hline
\textit{$\mu_{\mathrm{HARPSpre}} \mathrm{[m/s]}$} & $8.2_{-0.8}^{+0.8}$ & - & - & - \\
\textit{$\sigma_{w,\mathrm{HARPSpre}} \mathrm{[m/s]}$} & $0.66_{-0.12}^{+0.13}$ & - & - & - \\
\textit{$\mu_{\mathrm{HARPSpost}} \mathrm{[m/s]}$} & $-7.6_{-0.9}^{+0.9}$ & - & - & - \\
\textit{$\sigma_{w,\mathrm{HARPSpost}} \mathrm{[m/s]}$} & $0.17_{-0.15}^{+0.25}$ & - & - & - \\
\textit{$\mu_{\mathrm{HARPSpw}} \mathrm{[m/s]}$} & $1.9_{-1.1}^{+1.2}$ & - & - & - \\
\textit{$\sigma_{w,\mathrm{HARPSpw}} \mathrm{[m/s]}$} & $0.10_{-0.08}^{+0.40}$ & - & - & - \\
\textit{$\mu_{\mathrm{ESPRESSO}} \mathrm{[m/s]}$} & $9136.8_{-0.6}^{+0.7}$ & - & - & - \\
\textit{$\sigma_{w,\mathrm{ESPRESSO}} \mathrm{[m/s]}$} & $1.4_{-0.4}^{+0.5}$ & - & - & - \\
\hline
    \multicolumn{5}{c}{\textit{GP Parameters}} \\
    \hline
\textit{$\sigma_{\mathrm{GP,HARPSpre}} \mathrm{[m/s]}$} & $3.6_{-0.5}^{+0.5}$ & - & - & - \\
\textit{$\sigma_{\mathrm{GP,HARPSpost}} \mathrm{[m/s]}$} & $4.0_{-0.5}^{+0.6}$ & - & - & - \\
\textit{$\sigma_{\mathrm{GP,HARPSpw}} \mathrm{[m/s]}$} & $2.6_{-0.6}^{+0.8}$ & - & - & - \\
\textit{$\sigma_{\mathrm{GP,ESPRESSO}} \mathrm{[m/s]}$} & $0.16_{-0.14}^{+0.70}$ & - & - & - \\
\textit{$P_{\mathrm{rot,GP}} [\mathrm{days}^{-2}]$} & $38.6_{-0.5}^{+0.6}$ & - & - & - \\
\textit{$\Gamma_{\mathrm{GP}}$} & $4.2_{-0.7}^{+0.9}$ & - & - & - \\
\textit{$\alpha_{\mathrm{GP}} [\mathrm{days}^{-2}]$} & $0.00031_{-0.00010}^{+0.00017}$ & - & - & - \\
\hline\end{tabular}
\tablefoot{\tablefoottext{a}{Quasi-periodic kernel; derived parameters: (projected) planet mass $M \sin i$, the semi-major axis $a$, the equilibrium temperature $T_{\mathrm{eq}}$, stellar influx $S$, angular separation $\theta$ and contrast $C$ assuming an Earth-like albedo of 0.3.}}
\end{table*}

We computed a GLS periodogram of the residuals of model D to see if any periodic signals remain after subtracting the model. As can be seen in Figure \ref{fig:RVsignals} panel 7 another peak has appeared and reached a FAP of 0.0083 in the residuals. The period of this signal is very close to a 2:1 resonance with the $4.4~\mathrm{d}$ planet, but it becomes significant only after removing the signal of the $4.4~\mathrm{d}$ planet. Adding this candidate on an orbit with an eccentricity fixed at 0 to model D ($2.2~\mathrm{d}$ + D = model E) leads to an improvement in evidence of $\Delta\log\mathcal{Z}=4.42$ ($\Delta\log\mathcal{Z}=-3.27$ for the corrected evidences). The uncertainty on $\log\mathcal{Z}$ is $\sim 1$ for both models, which does not change much on the interpretation of the evidences. Allowing for an eccentricity in this model leads to a lower Bayesian evidence. We call this candidate GJ\,887\,f. The additional signal is fitted with a RV semi-amplitude of $K_f=0.37_{-0.08}^{+0.09}~\mathrm{m\,s^{-1}}$, namely, this is a $4.5\sigma$ detection. The ratio between the posterior periods of candidates e and f in the five-planet model is 1.99626$_{-0.00010}^{+0.00010}$:1.

An eccentric planet whose RV signal mimics the RV signal of a resonant pair \citep{Anglada-Escude2010} cannot be ruled out easily. 
A five-planet model with $2.2~\mathrm{d}$ and $4.4~\mathrm{d}$ planets on circular orbits is preferred by $\Delta\log\mathcal{Z}=5.93$ over an otherwise identical four-planet model ($4.4~\mathrm{d}$, $9.3~\mathrm{d}$, $21.8~\mathrm{d}$ and $50.8~\mathrm{d}$) with eccentric orbits for the inner three planets.
We also note that the evidence for the five-planet model E is a strong improvement over a four-planet model that does not include the $50~\mathrm{d}$ planet (E - $50~\mathrm{d}$ = model F). The Bayesian evidence of model E over model F is $\Delta\log\mathcal{Z}=9.42$ ($\Delta\log\mathcal{Z}=2.51$).
Posterior medians and confidence intervals of the five-planet model can be found in Table \ref{tab:results5p}.

\begin{figure}
    \centering
    \includegraphics[width=0.49\textwidth]{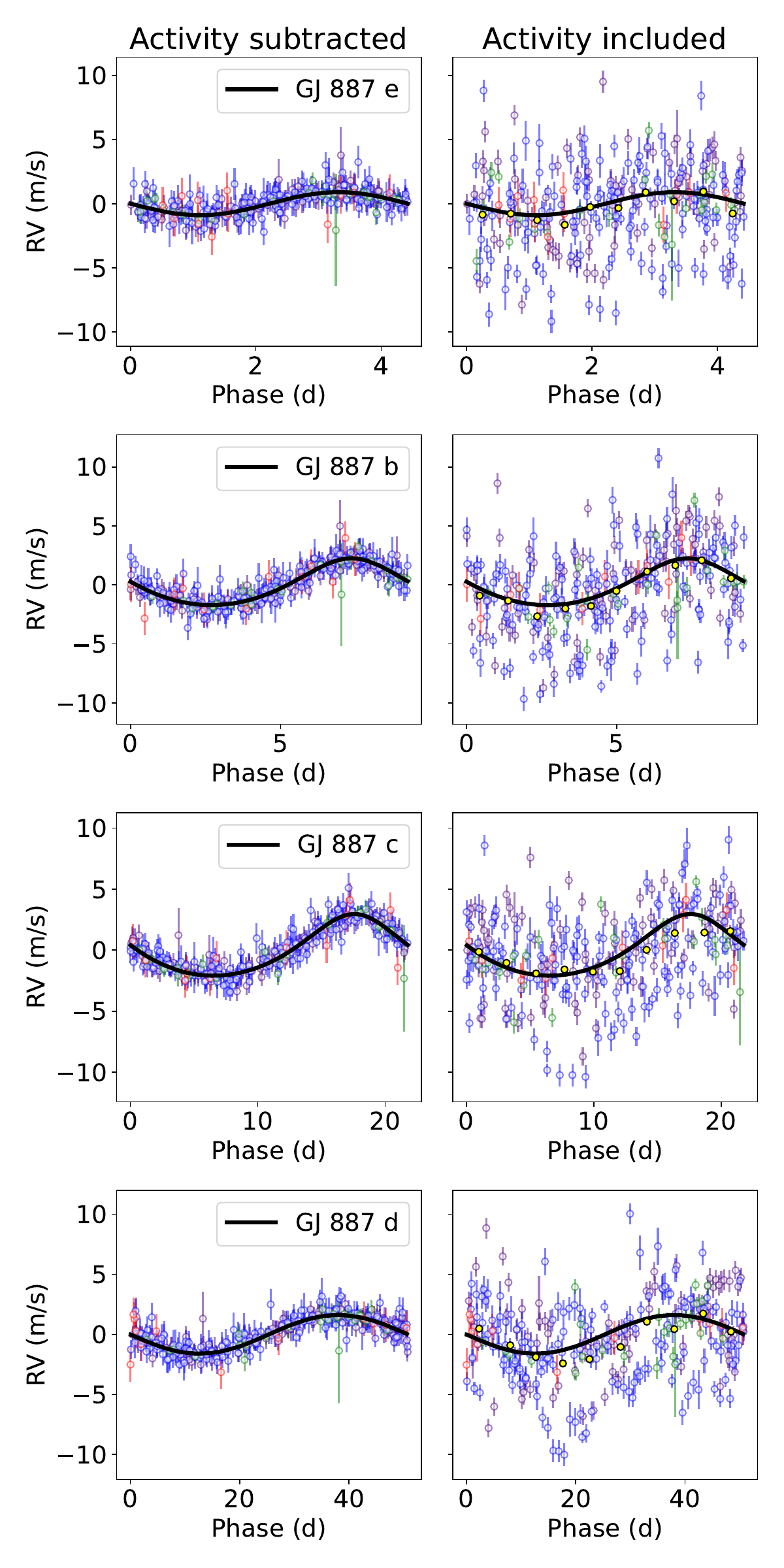}
    \caption{Phase folded plot of a four-planet model including candidates e, b, c, and d. The colors represent HARPSpre (purple), HARPSpost (blue), HARPSpw (green), and ESPRESSO (red) observations. In the left panels the model except for the respective planet component was subtracted from the RVs. In the right panels the model except for the planet and GP components are subtracted from the RVs. The yellow dots indicate phase-binned RV.}
    \label{fig:Phase}
\end{figure}

Since the addition of candidate f did not lead to an increase in evidence of $\Delta\log\mathcal{Z}>5$ over the model not including the signal, we report the four-planet model D as our preferred model (Table~\ref{tab:results4p}). Figure \ref{fig:Phase} shows the phase folded RV curves. The left panels are phase plots of the whole model: the fit is excellent with few residuals. To assess the impact of the GP, we consider the right panels, which display  the phase plots without subtracting the GP model from the RVs. All of the signals are recognizable in the binned phase plots; however, the RVs deviate significantly from the model. This shows that a GP is needed to account for stellar activity in order to detect the low-amplitude signals at $4.4~\mathrm{d}$ and $2.2~\mathrm{d}$ as the evidence for this signal only becomes significant for the GP model. The full GP model is given in \ref{fig:RVmodel}. The posterior distributions including correlation plots can be seen in Figures \ref{fig:cornerplanet} for the planet parameters, \ref{fig:cornerRV} for jitter and offset parameters, and \ref{fig:cornerGP} for GP parameters.

The planet parameters are all distributed as Gaussians, except for the arguments of periastron, $\omega$, for both eccentric planets. These show roughly uniform posteriors over the whole parameter space. The choice of an alternative parametrization with $\sqrt{e}\cos\omega$ and $\sqrt{e}\sin\omega,$ instead of $e$ and $\omega$ led to wider confidence intervals of the derived eccentricity and the overall Bayesian evidence was slightly worse so that we retained the standard parametrization of $e$ and $\omega$. 

The distributions of the jitters of HARPSpost and HARPSpw sets peak at $0~\mathrm{m\,s^{-1}}$. Other than that all parameters have distributions as expected. The $\alpha$ parameter seems to peak at 0, but this happens only due to the fact that we allowed the model to have high $\alpha$ parameters and, thus, a short correlation length. The distribution is expected to peak close to 0 in a linearly scaled plot.

The combination of a long observational base-line and dense sampling yields a high frequency resolution, so we could select very narrow priors for the periods of the inner planets. However, to test that the low-amplitude signals for candidates e and f are not dependent on the narrow priors, we re-ran models D and E with wider priors for the periods of the shallow signals. In model D, the prior of the period of candidate e (f) was $\mathcal{U}(4.42,4.43)~\mathrm{d}$ ($\mathcal{U}(2.21,2.22)~\mathrm{d}$), namely, the prior spans $0.01~\mathrm{d}$. The recovered posterior period was $P_e=4.42490_{-0.00013}^{+0.00013}~\mathrm{d}$ ($P_f=2.21662_{-0.00010}^{+0.00010}~\mathrm{d}$), so the $1\sigma$ confidence interval is only $2.6\%$ ($2.0\%$) of the prior. Widening the prior by a factor of 10 (15) to $\mathcal{U}(4.4,4.5)~\mathrm{d}$ ($\mathcal{U}(2.15,2.3)~\mathrm{d}$) yields essentially the same posterior period: $P_e=4.42490_{-0.00013}^{+0.00013}~\mathrm{d}$ ($P_f=2.21656_{-0.00011}^{+0.00011}~\mathrm{d}$). The posteriors are not significantly affected by the prior choice. 

To confirm that the results for the planet configurations and parameters do not depend on the activity model, we ran all models with a dSHO kernel. The overall results are similar to the QP models: the rotation period was around $39~\mathrm{d}$ and the planet parameters agree within $1-2\sigma$. Additionally, the models including the $4.4~\mathrm{d}$ planet perform much better than those that do not.
There is moderate evidence for the $2.2~\mathrm{d}$ candidate, as there was in the QP model.
In the dSHO models however, the evidence for the $50~\mathrm{d}$ planet decreases as the difference between Bayesian evidence between models C$_{\rm dSHO}$ and D$_{\rm dSHO}$ is only $\Delta\log\mathcal{Z}=6.05$. When including the $2.2~\mathrm{d}$ planet the difference between the models is $\Delta\log\mathcal{Z}=3.31$ (E$_{\rm dSHO}$ over F$_{\rm dSHO}$). 
This could be because the dSHO model has more parameters and is therefore more flexible. The $50~\mathrm{d}$ signal could be partly absorbed by the GP, which is an effect that happens more efficiently with a more flexible model.
The Bayesian evidences of all models we ran are shown in Table \ref{tab:results}. 

For the posteriors of the models D and E we tested the (re-sampled) posterior configurations for stability with SPOCK.
Some of the samples with higher eccentricities of planets b and c were down-weighted, however, neither of the median values of the posterior distributions was changed significantly by factoring in the stability. The adjusted medians and confidence intervals of the weighted posteriors are tabulated in Tables \ref{tab:results4p} and \ref{tab:results5p}.

We also computed the mutual Hill separation \citep{Weiss2018}, with the separation in units of mutual Hill radii, 
\begin{equation}
    \Delta=2\frac{a_{j+1}-a_{j}}{a_{j+1}+a_{j}}\left(\frac{3M}{m_{j+1}m_j}\right)^{1/3}
,\end{equation}
for neighboring pairs of planets $j$ and $j+1$ with the semi-major axes, $a$, the masses, $m,$ and the host star mass, $M$. For the five-planet configuration febcd in this system we obtain mutual Hill separations of $28.6^{+0.9}_{-1.0}$ (fe), $21.6^{+0.6}_{-0.6}$ (eb), $20.0^{+0.6}_{-0.6}$ (bc), and $18.6^{+0.9}_{-0.8}$ (cd). For other planet configurations these separations vary only insignificantly. These values are typical for multiplanet systems \citep{Weiss2018}.

To further examine the short period signals we re-ran models A - F with the unbinned data. This could allow a better representation of the RV signals. The evidence for model E over model D however, was only $\Delta\log\mathcal{Z}=1.78$ ($\Delta\log\mathcal{Z}=-5.91$ corrected), which could be evidence against the $2.2~\mathrm{d}$ candidate compared to the QP model, but it could also be the effect of short-term activity, which is the reason the data were binned in the first place.

\subsection{Injection-recovery tests}

To show that planet signals can be recovered and confirmed by Bayesian model comparison as it is used in section \ref{sec:modelRVs} in the GP models, we run injection-recovery tests similar to those of \cite{Rajpaul2024}. We are particularly interested in the recovery of low-amplitude short-period signals similar to the candidates e and f and periods above the rotation period similar to candidate d. Therefore, we added sinusoidal signals on the original RVs. We selected periods similar to the previously mentioned candidates in regions of the periodogram that show no other significant peaks so that the injected signals are not confused with potential additional signals in the data. We selected the periods $3.1~\mathrm{d}$ as short-period injection and $66~\mathrm{d}$ as long-period injection and injected two different RV semi-amplitude signals in each of these periods. As semi-amplitudes for the short-period signal we used $K_{inj}=0.94~\mathrm{m\,s^{-1}}$ and $K_{inj}=0.37~\mathrm{m\,s^{-1}}$ just as the recovered signals from the candidates e and f. For the long-period signal we injected $K_{inj}=1.8~\mathrm{m\,s^{-1}}$, as recovered for candidate d, and to test the capability of the model, we injected $K_{inj}=1.2~\mathrm{m\,s^{-1}}$ at the same period. With this procedure we created four new time series and for each of these we ran a four-planet model (D) and a five-planet model with a prior including the period of the injected signal (D + injected signal). We then compared the Bayesian evidence of these two models for each time series to determine if the recovery was successful. The widths of the (uniform) priors of the injected planets were similar to the corresponding planets f, e, and d and are tabulated in Table \ref{tab:priorsinj}.

For the short-period signal we recovered the $K_{inj}=0.94~\mathrm{m\,s^{-1}}$ with $\Delta\log\mathcal{Z}=29.20$, which is a clear detection and the results are similar to the detection of candidate e. The recovered amplitude was $K_{rec}=0.88_{-0.08}^{+0.08}~\mathrm{m\,s^{-1}}$. For the injected $K_{inj}=0.37~\mathrm{m\,s^{-1}}$ the five-planet model performs better than the four-planet model by $\Delta\log\mathcal{Z}=5.19$ and the recovered amplitude was $K_{rec}=0.43_{-0.09}^{+0.09}~\mathrm{m\,s^{-1}}$. This suggests a strong detection. The difference in Bayesian evidence is comparable to the difference we saw between the four-planet and five-planet models in the original data. That parallel shows the model comparison is behaving consistently: it reacts the same way both to the injected signal we know is real and to the $2.2~\mathrm{d}$ candidate in the actual data.
We note that the recovered period of the injected planet was $P_{rec}=3.10368^{+0.00018}_{-0.00019}~\mathrm{d}$ which is significantly different from the injected $P_{inj}=3.1~\mathrm{d}$, which is a sampling effect caused by timing of observing periods as one large chunk of observations  was recorded in 2005 and another large chunk was recorded in 2019, which leads to a sampling frequency of $\sim\frac{1}{5000~\mathrm{d}}$ which is consistent with the aliasing pattern. Similar patterns can be observed when zooming in on the periods of the other signals, which can be seen in Figure \ref{fig:sbgls}. 
For this experiment it does not play a role as we were interested in the Bayesian model comparison of the models.

For the long-period injection we were able to recover the injected planet with $K_{inj}=1.8~\mathrm{m/s}$ with an evidence $\Delta\mathrm{log}\mathcal{Z}=11.53$ and a recovered amplitude of $K_{rec}=1.75_{-0.24}^{+0.26}~\mathrm{m/s}$, however, for the $K_{inj}=1.2~\mathrm{m/s}$ signal, we only recovered the signal with an evidence of $\Delta\log\mathcal{Z}=1.67$ with an amplitude of $K_{rec}=1.30_{-0.28}^{+0.4}~\mathrm{m/s}$ over the four-planet model. This shows that a GP can interfere with planet signals close to the period of the GP.

In conclusion, in the idealized conditions of a circular orbit, with no dynamical interaction with other planets, and in a low noise region of the GLS periodogram, the signal could just be detected. For the candidate f whose orbit could be eccentric or affected by dynamical interactions over time this implies a detection of a planet with the recovered parameters is within reach, but it is on the edge of detectability with the current data. The Bayesian model comparison values for the injected signals match those obtained for the real planets and candidates when using similar priors, which validates our method of using narrow priors.

\subsection{Detection limits}

\begin{figure}
    \centering
    \includegraphics[width=0.49\textwidth]{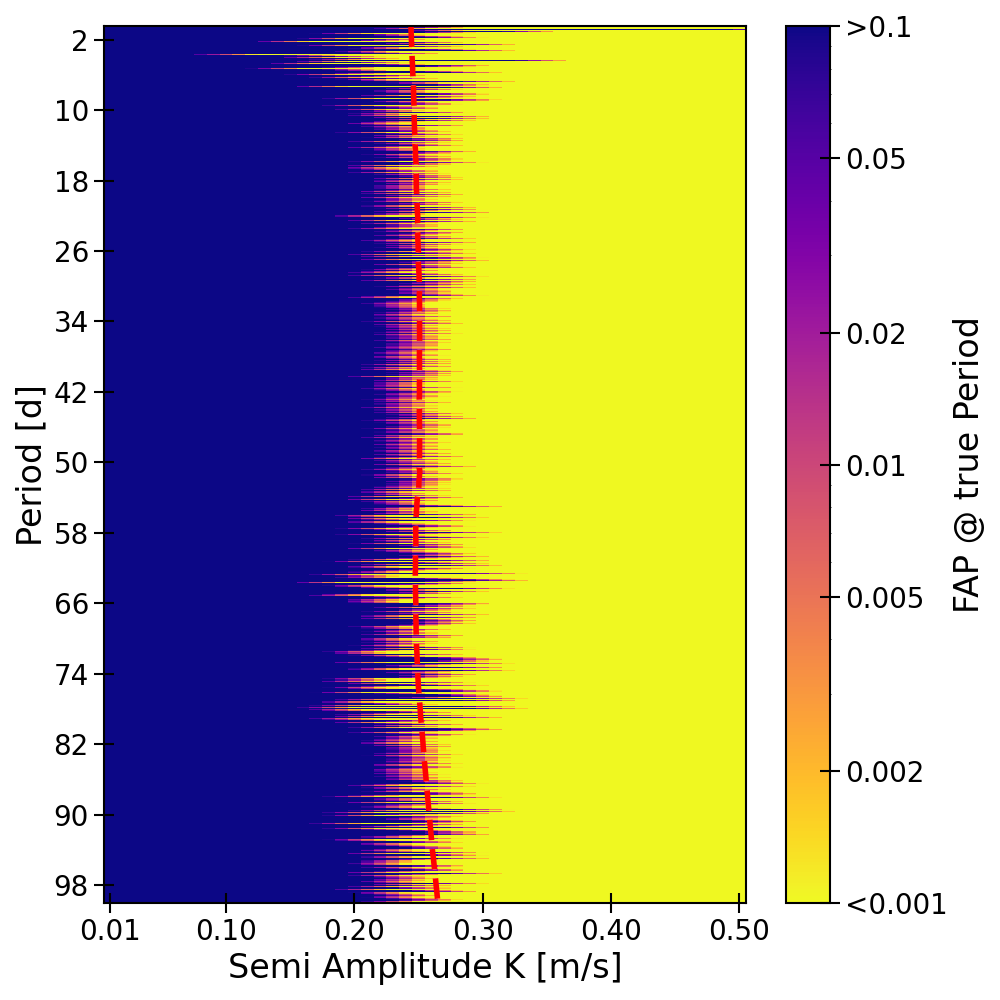}
    \caption{FAP of injected signal with given period and amplitude. The red dashed line describes the smoothed 0.01 FAP level.}
    \label{fig:detlimits}
\end{figure}

We can describe detection limits in a more formal way with the methodology of \citet{Liebing2024}.  The estimation of the detection limit works under the assumption that no signals are present in the residuals and only noise remains. Because the $2.2~\mathrm{d}$ signal remains in the four-planet model residuals, we use the five-planet model residuals. We add a Keplerian signal of a noneccentric planet to the RV residuals of the five-planet GP model. Then we use the FAP calculated by the GLS implementation from \citet{Zechmeister2009} at the true, injected period to determine the detectability of the signal. This is done for a grid of amplitudes $K$ and periods $P$. Our grid spans $0.5-100~\mathrm{d}$ with a step size of $0.1~\mathrm{d}$ and $0.1-1~\mathrm{m\,s^{-1}}$ with a step size of $0.01~\mathrm{m\,s^{-1}}$. We determine the detection threshold using the value of $K$ below which the signal has FAP $ > 0.01$. Figure~\ref{fig:detlimits} shows the FAP values over the amplitude-period grid. The red dashed line is the threshold, smoothed by a Savitzky-Golay filter. The mean detection limit over $0.5-100~\mathrm{d}$ is $25\pm3~\mathrm{cm\,s^{-1}}$. For short-period signals around $2~\mathrm{d}$ the estimated detection limit is around $23~\mathrm{cm\,s^{-1}}$, which would allow the detection of a $37~\mathrm{cm\,s^{-1}}$ signal, such as candidate GJ 887 f. For longer periods ($>20~\mathrm{d}$) the detection limit increases to $\sim 26~\mathrm{cm\,s^{-1}}$. However, as shown in the previous section, injected signals interact with the GP and the residuals are affected by the GP.

\subsection{Coherence of signals}

To test whether the candidate planet signals are coherent, we computed sBGLS periodograms for the relevant period ranges. The sBGLS tracks how the significance of a periodicity evolves as data are added sequentially, enabling incoherent activity-induced signals to be distinguished from stable planetary ones.
For the stronger signals of planets b, c, d, and the stellar rotation period, we pre-whiten the data by removing all planetary signals and the rotation period as a Keplerian, except for the one being tested. For the signals of candidates f and e we use the residuals of the GP models including planets b, c and d (and e for the periodogram of planet f).\newline

In Figure \ref{fig:sbgls} we see clearly coherent signals for the candidates f, e, and b. For candidates c and d the signal is most evident in the end of observations, which is characteristic for a coherent signal, however both signals peak around 150-170 and drop off before peaking in the end. This first peak appears for both signals approximately at the time where the star becomes active. Consequently the RVs are dominated by the stellar activity there and interfere with the signals of the candidates. In the end when the activity drops becomes weaker according to the indicators, the candidate signals become stronger again. For the rotation period at $39~\mathrm{d}$ the signal appears coherent at first sight. However, a closer look reveals a drop-off for the last observations. This happens, because the star becomes less active at this period of time.

\begin{figure*}
    \centering
    \includegraphics[width=0.99\textwidth]{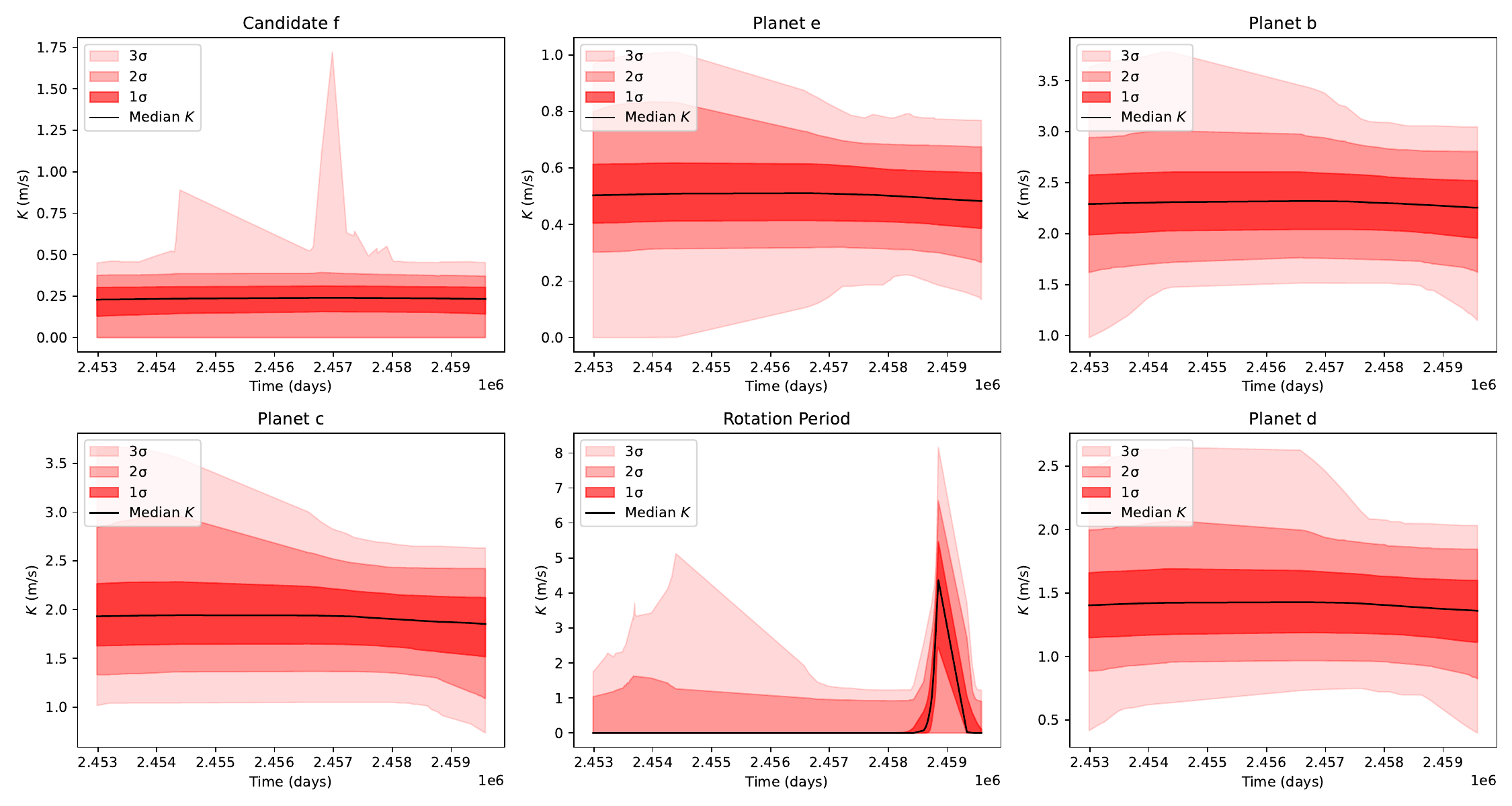}
    \caption{Stability test with apodised models for the candidate signals, showing the evolution of the effective semi-amplitude over the observational baseline.}
    \label{fig:apodised}
\end{figure*}

As a second approach, we investigated in the same RV residuals as described for sBGLS, the temporal stability of the detected periodicities using apodised signals models \citep{Gregory2016,Hara2022b}, which introduce a Gaussian envelope to track how the signal amplitude evolves over time. This technique, employed by \citet{SuarezMascareno2025} to validate the coherence of planetary signals, enables us to distinguish genuinely stable, planet-like periodicities from activity-induced signals. As test periods we used posterior medians as fixed values from Table \ref{tab:results5p}. The parameters of the Gaussian from \citet{SuarezMascareno2025} and remaining planet parameters are fitted using MCMC. All planets and candidates have an amplitude that is stable in the median and within 1 and 2 $\sigma$ intervals (Figure \ref{fig:apodised}). Only for candidate f the $3~\sigma$ confidence interval, shows a stronger signal around $2,457,000$ BJD. We note that the amplitude of the signal discussed here is not necessarily equal to the amplitudes inferred from the GP models. In this analysis, each signal is fitted individually after removing the contributions of other planets and stellar activity, following the procedure described in the sBGLS analysis. The rotation period in contrast shows a signal that is very strong around around 2,459,000 BJD despite not being present at all in the median before that. This shows that the signal of the rotation period at $38.7~\mathrm{d}$ is incoherent.

\subsection{Transit search}
\citet{Jeffers2020} detected no planet transits in TESS Sector 2. With two additional TESS Sectors (28 and 69) we revisited this question, searching for transits of the planets and candidates we found in the RVs. We flattened the light curves using splines with a window length of $1~\mathrm{d}$ from the \texttt{W{\={o}}tan} package \citep{Hippke2019b} (upper three panels of \ref{fig:Tsearch}). To find potential transits we applied the TLS algorithm. In the lower panel of Figure \ref{fig:Tsearch} we can see the SDE of the combined light curve. None of the peaks are significant; no peaks coincide with the RV periods. We can estimate the radius of the planets from their minimum masses from probabilistic models of \citet{Chen2017}. The radii $R_{\rm f}=0.81~\mathrm{R}_\oplus$, $R_{\rm e}=1.12~\mathrm{R}_\oplus$, $R_{\rm b}=1.77~\mathrm{R}_\oplus$, $R_{\rm c}=2.39~\mathrm{R}_\oplus$, and $R_{\rm d}=2.29~\mathrm{R}_\oplus$ combined with the stellar radius of $R_{\rm s}=0.468~\mathrm{R}_\oplus$ predict transit depths of $D_{\rm f}=0.00025$, $D_{\rm e}=0.00048$, $D_{\rm b}=0.00120$, $D_{\rm c}=0.00219$, and $D_{\rm d}=0.00201$. The median error of the normalized TESS flux is 0.00011 so that transits of each candidate would be detectable.
Although the limited temporal coverage of TESS leaves open the possibility of missed events, no transit signatures are detected in the light curves. The inner planets are therefore non-transiting, and, given both the absence of signals and the lower geometric transit probability of outer planets, it is unlikely that the outer planets transit.

\section{Discussion}\label{Discussion}
We 
determined the rotation period of this important nearby star, and significantly improved 
the RV analysis. 
We find strong evidence for 
planet GJ 887 d in the HZ.  We discuss our results below. 

\subsection{Rotation period}\label{sec:Protdisc}
\citet{Jeffers2020} suggested rotation periods of around $37~\mathrm{d}$ and $55~\mathrm{d}$, but their study was hampered because
the star was inactive during their observations.
Our new data include coverage during a time when GJ\,887 was more active. 

We estimated the rotation period of GJ 887 in Section~\ref{sec:Prot}.
ASAS photometry hints at a period around $39~\mathrm{d}$. Several activity indicators: FWHM, BIS, dLW, H$_\alpha$, Na D$_1$, and Na D$_2$, as well as the RVs also reveal a $39~\mathrm{d}$ periodicity. In dLW, H$_\alpha$, Na D$_1$, and Na D$_2$, we see a weaker signal at $49~\mathrm{d}$; however, since it does not show up in photometry and several indicators it is unlikely to be the rotation period.

In 2019 many spectra were obtained on consecutive nights, and cover a strong activity outbreak. Measuring times between the peaks in the time series yields $39~\mathrm{d}$ for the RV, dLW, and Na D$_1$ time series. In the analysis using apodised models the signal shows a strong amplitude of several $\mathrm{m\,s^{-1}}$ so that the signal is incoherent and therefore not a planet. Lastly, all GP models with high Bayesian evidence suggest a rotation period around $39~\mathrm{d}$. Our preferred model, the four-planet model, model D, suggests a rotation period of $P_{\mathrm{rot}}= 38.7_{-0.5}^{+0.5}\mathrm{days}$ when searching with a wide prior of $\mathcal{J}(10,100)$. When modeling the rotation period with a Keplerian, residuals appear at $P_{rot}/2$, which is typical for the rotation period. The signal around $50~\mathrm{d}$ which \citet{Jeffers2020} attributed to either stellar activity or a planet, we now interpret as a planet.

\subsection{Assessment of signals in RVs}

We found multiple significant signals in the GLS periodograms when subtracting significant signals iteratively at periods $9~\mathrm{d}$, $21~\mathrm{d}$, $50~\mathrm{d}$, $4.4~\mathrm{d,}$ and $2.2~\mathrm{d}$.\newline

The strongest signals are at $9~\mathrm{d}$ and $21~\mathrm{d}$ and belong to the previously reported planets b and c \citep{Jeffers2020}. The FAPs in the GLS periodograms are $5.8\times10^{-8}$ and $8.0\times10^{-9}$ and respectively $1.4\times10^{-10}$ and $2.0\times10^{-10}$ in the $\ell_1$ periodogram. Fitting both signals in Bayesian models treating the stellar rotation as GP improves the Bayesian evidence by $\Delta\log Z=34.56$ and $\Delta\log Z=14.08$. Correcting the evidences for the choice of a narrow prior still leaves strong evidences for both signals ($\Delta\log Z=26.82$ and $\Delta\log Z=7.50$).
The $9~\mathrm{d}$ signal is perfectly coherent in the sBGLS periodogram. The $21~\mathrm{d}$ signal does not appear perfectly coherent, however the incoherence appears in the active phase of the star, where the RV signal is dominated by stellar activity. The amplitude of the signal appears stable over the complete time series according to analysis using apodised models. Consequently, we reported both signals as planets, in agreement with \citet{Jeffers2020} . 

The signal at $50~\mathrm{d}$ shows a FAP of $1.6\times10^{-6}$ in the GLS periodogram after subtracting planets b and c and the rotation period with Keplerians and FAP of $5.2\times10^{-3}$ in the $\ell_1$ periodogram. Including the $50~\mathrm{d}$ planet in GP models
improves the evidence in all models with $\Delta\log\mathcal{Z}=6.48$ when not fitting the $2.2~\mathrm{d}$ signal and by $\Delta\log\mathcal{Z}=9.57$ when including the signal in the model. When correcting for the choice of a narrow prior of the periods, this evidence decreases to $\Delta\log\mathcal{Z}=-0.43$ and $\Delta\log\mathcal{Z}=2.51, $  thereby offering only moderate evidence for the signal at best. Using the corrected evidence for the $50~\mathrm{d}$ signal from the non-GP models, we obtained strong evidence for the existence of the planet ($\Delta\log\mathcal{Z}=12.29$).

In the four-planet GP model, we reached a $4.6\sigma$ detection considering the semi-amplitude and its uncertainty ($K_d=1.7_{-0.4}^{+0.4}$). The signal is consequently significant. The signal shows a similar pattern in the sBGLS periodogram as the confirmed $21~\mathrm{d}$ planet. The amplitude of the planet remains stable for the whole time series using apodised models. The Na D indicators and the dLW show activity at a period of $48~\mathrm{d}$, however, most of the indicators do not show signals around $50~\mathrm{d}$. Additionally, the semi-amplitude of the signal in the posterior remains stable independent of how the stellar activity is dealt with. Therefore, we conclude that the signal is most likely caused by a planet and we confirm the signal as planet d.

The signal at $4.4~\mathrm{d}$ is not significant when treating stellar activity with a Keplerian. However, it shows a FAP of $4.6\times10^{-11}$ in the residuals of the three-planet GP model. In the $\ell_1$ periodogram the signal is detected with a FAP of $2.6\times10^{-4}$. Including the planet in a GP model improves the Bayesian evidence by $\Delta\log\mathcal{Z}=27.15$ ($\Delta\log\mathcal{Z}=18.22$ for corrected evidences) compared to the three planet model. The uncertainty on the semi-amplitude $K_e=0.91_{-0.10}^{+0.10}~\mathrm{m/s}$ suggests a $9.8\sigma$ detection. There is no signal at this period in the activity indicators and the signal is coherent in the sBGLS periodogram. The analysis using apodised models shows a constant period over the whole period of observations. Consequently, this is a clear detection of a new planet that we call GJ 887 e.

\begin{figure}
    \centering
    \includegraphics[width=0.49\textwidth]{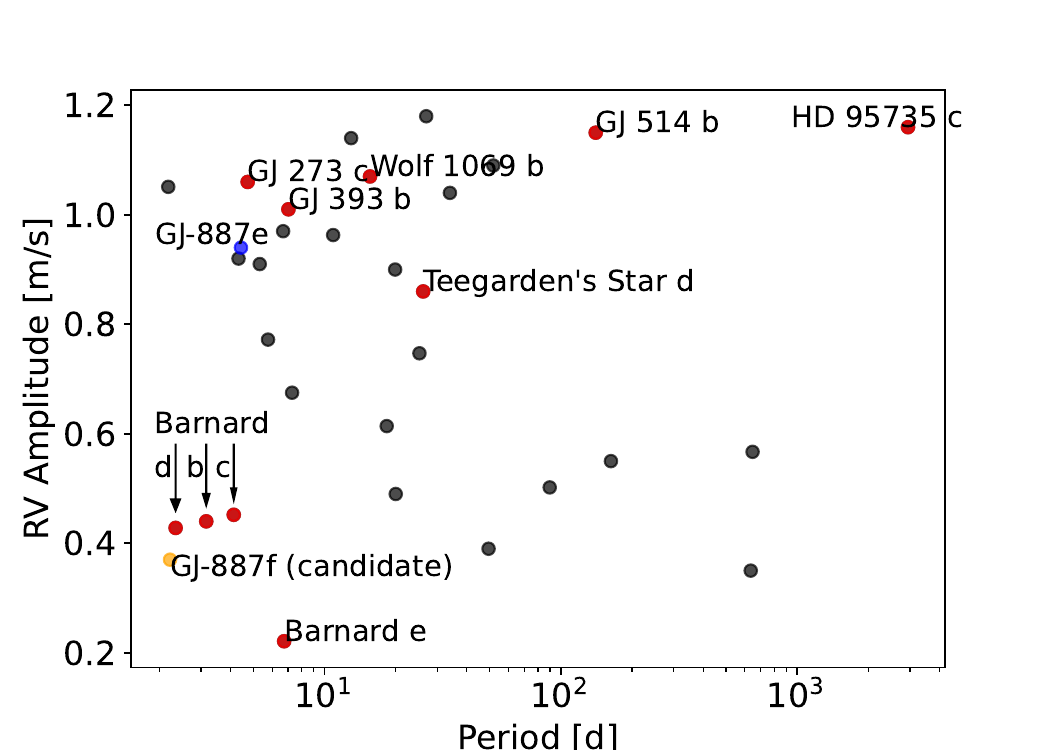}
    \caption{Planets from the NASA Exoplanet Archive with an RV-amplitude below $1.2~\mathrm{m/s}$ that were first discovered with the RV method and that do not have a controversial flag in the archive. The planets are plotted period against RV amplitude. Planets around M stars are highlighted in red and GJ 887 e and f are highlighted in blue and orange.}
    \label{fig:lowmassplanets}
\end{figure}

The signal of the $2.2~\mathrm{d}$ candidate is neither significant in the $\ell_1$-periodogram nor in the GLS periodogram without subtracting a GP model. However, when subtracting the four-planet GP model including the planets confirmed above, the residuals show a signal with a FAP of $8.3\times10^{-3}$ at $2.2~\mathrm{d}$. Adding the $2.2~\mathrm{d}$ signal improves the fit by $\Delta\log\mathcal{Z}=4.42$ compared to the four-planet model, however, when correcting the evidence for the choice of prior the evidence results in a negative Bayes evidence ($\Delta\log\mathcal{Z}=-3.27$). In the five-planet model, the signal yields a $4.5~\sigma$ detection of the semi-amplitude, $K_f=0.37_{-0.09}^{+0.08}~\mathrm{m/s}$. According to common standards a model needs to show a significance of $\Delta\log\mathcal{Z}>5$ for being considered a planet, which was not achieved in this model. Still, with additional RVs from an ultra-precise instrument such as ESPRESSO, confirmation could be possible if the signal is indeed planetary in origin.
Activity plays no role on the timescale of the signal for this star and the signal is coherent in the sBGLS periodogram and in the analysis using apodised models. An alternative explanation is that the signal arises from dynamical interactions with other planets, leaving a residual from the $4.4~\mathrm{d}$ planet. This is plausible since the $2.2~\mathrm{d}$ feature lies close to a mean-motion resonance with planet e. The posterior distribution of the resonant angle 
$\phi = 2\lambda_{2} - \lambda_{1}$ 
is strongly clustered (resultant length $R = 0.84$), 
peaking near $\pi/2$ with an approximate dispersion 
$\sigma \approx 0.58~\mathrm{rad}$. 
This clustering indicates the angle is likely librating 
rather than circulating, which is suggestive of a 2{:}1 
mean-motion resonance.

\citet{Harada2025} investigated the system using 262 archival HIRES and HARPS RVs. Our dataset, comprising 277 HARPS and 12 ESPRESSO RVs, is bigger and consequently more sensitive. The cited authors used less HARPS data and did not treat the rotation period with a GP model; therefore, the $4.4~\mathrm{d}$ and $50~\mathrm{d}$ planets were not detected in that work.

\begin{figure*}
    \sidecaption
    \centering
    \includegraphics[width=12cm]{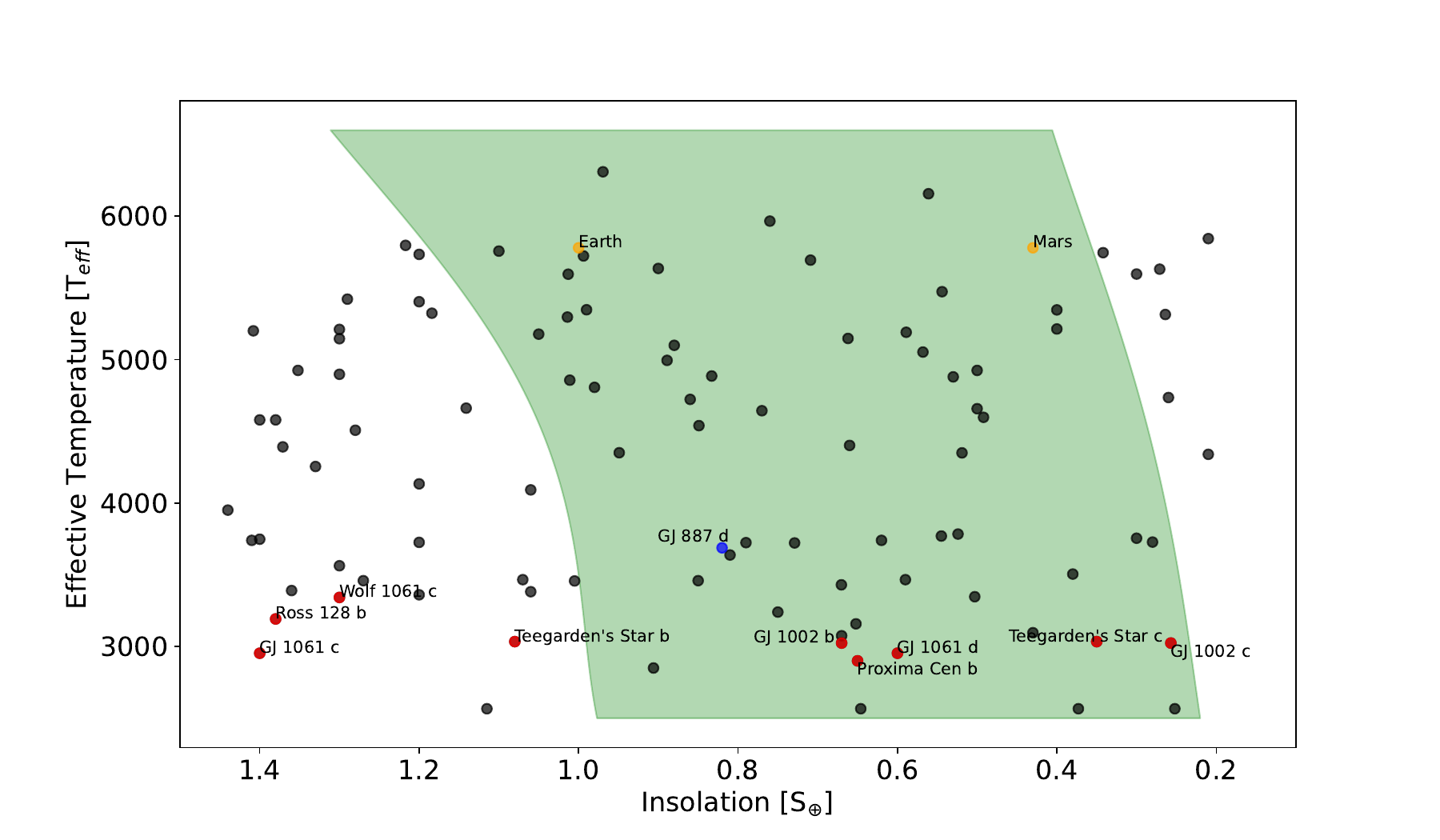}
    \caption{Planets in the NASA Exoplanet Archive with their effective temperatures against their insolation. The HZ (green) according to \citet{Kopparapu2014} for a $5~\mathrm{M}_{\oplus}$ planet. Close-by planets ($d<5~\mathrm{pc}$) are highlighted in red. GJ 887 d is marked blue. Solar system planets are marked in orange.}
    \label{fig:hz}
\end{figure*}

\subsection{Detectability of RV signals}

To investigate the impact of the GPs on the detectability of planets in our data, we performed injection-recovery tests for long and short period planets. The injected RV signals correspond to the idealized situation of a planet on a circular orbit that does not interact with the other planets, with a period unaffected by aliases or harmonics of other signals. 

The injected $66~\mathrm{d}$ signal with the same amplitude as GJ 887 d was easily recovered. However, when decreasing the amplitude to $1.2~\mathrm{m\,s^{-1}}$, the signal could no longer be detected. This shows the danger of a GP that acts on a timescale similar to the periods of the planets. A shallow planetary signal can be absorbed or weakened by the GP and can therefore remain undetected.

The short-period planet with $K_{inj}=0.94~\mathrm{m\,s^{-1}}$ could be recovered easily, with a similar improvement in $\Delta\log\mathcal{Z}$ as planet e. For the $K_{inj}=0.37~\mathrm{m\,s^{-1}}$ injection we obtained a similar evidence as for candidate f, however, the an alias of the injected period was recovered. We conclude that with the current dataset and our model we are not quite able to find strong evidence for a planet of this amplitude at this period.

Figure \ref{fig:lowmassplanets} shows planets that were first detected by RVs. 
Few planets with $K < 1~\mathrm{m\,s^{-1}}$ have been detected around M stars: only the recently detected Teegarden's\,Star\,b \citep{Dreizler2024}, Barnard\,b \citep{GonzalesHernandez2024} and Barnard\,cde \citep{Basant2025}; both orbit inactive stars. Teegarden's\,Star\,b was confirmed without a GP model. \citet{GonzalesHernandez2024} confirm a short-period signal with a $55~\mathrm{cm\,s^{-1}}$ amplitude (Barnard\,b) and three short-period candidates with amplitudes of $20-47~\mathrm{cm\,s^{-1}}$. With more data \citet{Basant2025} confirmed these three candidates modeling stellar activity with a univariate GP, similar to the analysis that has been carried out in this work. GJ\,393\,b \citep{Amado2021} has a similar amplitude in a similar orbit to GJ\,887\,e and was confirmed using a GP. Around K and G stars, a number of sub-meter-per-second ($\mathrm{m\,s^{-1}}$) detections have been made. Notably, the systems tau Cet \citep{Feng2017} and HD\,215152 \cite{Delisle2018} host multiple planets with sub-$\mathrm{m\,s^{-1}}$ amplitudes first detected by RVs. All of these systems were observed several hundred times by high-precision spectrographs (e.g., ESPRESSO, HARPS, CARMENES) with a sufficient cadence to sample the periods of the candidates.

Our injection-recovery tests showed that even lower amplitude planets could be found in these short-period orbits with a FAP of 0.01. The detection of planet GJ\,887\,e and the potential detection of candidate GJ\,887\,f are part of ongoing progress to detection of lower RV amplitudes. This will help improve our understanding of the properties of low-mass planets.

\subsection{Habitability of GJ\,887\,d}

Planet d with $P = 50.74^{+0.04}_{-0.04}$ has instellation of $0.81^{+0.13}_{-0.12}~\mathrm{S}_{\oplus}$ using the posteriors of the planetary parameters and the reported Gaussian errors on the stellar parameters (Table \ref{tab:stellarparams}). The planet is in the HZ \citep{Kopparapu2014} of its host star (Figure~\ref{fig:hz}). The HZ for GJ 887 spans from planetary orbits of 43 to 122 days. With a minimum mass of $M\sin i=6.5^{+1.5}_{-1.5}~\mathrm{M}_{\oplus}$ GJ 887 d is a super-Earth ($2~\mathrm{M}_{\oplus}<M<10~\mathrm{M}_{\oplus}$) for inclinations $i>41^\circ$. Without an independent radius estimate the density and hence the composition of the planet cannot be determined. According to \citet{Luque2022} planets in this mass range have either a rocky, a water-world or a puffy sub-Neptune composition.

In the coming decades an atmospheric characterization of GJ\,887\,d might be possible with direct imaging missions from space. Due to its brightness and its distance of only $3.29~\mathrm{pc}$ from the Sun, GJ\,887 is on the target list of the proposed missions: LIFE \citep{LIFE}, the Habitable Exoplanet Observatory (HabEx; \citet{HabEx}) and HWO \citep{Mamajek2024}. GJ\,887\,d could therefore be an ideal target to detect biomarkers on a HZ planet.\newline

For our target planet, we find an angular separation of $64.6_{-2.2}^{+2.1}~\rm mas$ and a reflected-light contrast of $38.1_{-2.4}^{+2.8}\times10^{-9}$, assuming an Earth-like albedo of 0.3. HWO is expected to achieve an inner working angle (IWA) of $65~\rm mas$ (Tier~C) \citep{Mamajek2024}. The mission’s contrast floor is set near $(2.5{-}4)\times 10^{-11}$ \citep{Mamajek2024}. Thus, the planet's angular separation from the star is on the edge of the inner working angle making detectability with HWO unclear. Its brightness ratio is orders of magnitude above the HWO detection threshold making the angular separation the dominant limitation.\newline

GJ\,887 displays a low level of magnetic activity.
\cite{Mesquitaetal2022MNRAS} estimated  GJ887's HZ
to be in Earth-like stellar wind conditions and to receive an Earth-like level of Galactic cosmic rays.
However, GJ\,887 can nonetheless exhibit some flaring activity \citep{Loyd2020RNAAS} which can in principle affect the habitability of GJ\,887\,d.
The interactions of stellar wind, flares, and cosmic rays with GJ\,887\,d are strongly dependent on the planetary magnetic field, which remains so far unknown.

\section{Summary}\label{Summary}
Our analysis of 277 HARPS and 12 ESPRESSO RVs recovered \citet{Jeffers2020}'s two reported planets. We confirmed a planet in a $50~\mathrm{d}$ orbit within the HZ that was detected by \citet{Jeffers2020}. Additionally, we detected and confirmed a planet in a $4.4~\mathrm{d}$ orbit with a sub-$\mathrm{m\,s^{-1}}$ RV amplitude. A further coherent signal at $2.2~\mathrm{d}$ in the residuals of the four-planet GP model is close to 2:1 resonance with the $4.4~\mathrm{d}$ planet, with an amplitude of $K = 0.37^{+0.09}_{-0.09}~\mathrm{m\,s^{-1}}$. There is evidence that this signal belongs to a planet; however, the statistical significance is insufficient to claim this signal a planet. The amplitude of the signal is below the stability excursions of the HARPS spectrograph. 
Future investigations with high-precision spectrographs could help to identify the origin of the signal. Our signal detections were made possible by new RedDots observations with a daily cadence. These were ideal to sample the periods of the planets and captured two rotations of the star, with the period $P_{rot} \sim 39~\mathrm{d}$ visible in several activity indicators and the RVs. The same rotation period was found in ASAS photometry and  also recovered by the GP modeling of the stellar activity.

GJ\,887 is a compelling system for further study. It is a nearby and, hence, bright, M dwarf, hosting a minimum of four planets including a super-Earth-mass, Earth-mass, and potentially sub-Earth-mass planets. At least one of the planets is in the habitable zone.

\begin{acknowledgements}
CH and ACC acknowledge support from UKRI/ERC Synergy Grant EP/Z000181/1 (REVEAL). JRB and CAH were funded by the Science and Technology Facilities Council under consolidated grant ST/T000295/1 and ST/X001164/1. FDS acknowledges support from a Marie Curie Action of the European Union (Grant agreement 101030103). We thank the anonymous referee for their careful reading of the manuscript and for their constructive comments, which helped improve the clarity and quality of the paper.
\end{acknowledgements}

\bibliographystyle{aa} 
\bibliography{lib} 

\begin{appendix}
    
\section{Supplementary information}

\begin{table}[h!]
    
    \centering
    \caption{Chronological 
    list of observations (before binning) giving PI, number of observations and Program ID\tablefootmark{a}.  }
    \begin{tabular}{llll}
        \hline
        \textbf{Dataset} & \textbf{PI} & \textbf{$N_{obs}$} &\textbf{Program ID} \\
        \hline\hline
        HARPSpre                   & Mayor& 78  & 072.C-0488(E)\tablefootmark{b}\\
        HARPSpre                   & Lagrange&  28     & 192.C-0224(A)\tablefootmark{b}\\
        HARPSpre                   & Lagrange& 2      & 192.C-0224(B)\tablefootmark{b}\\
        HARPSpre/post              & Lagrange&14& 192.C-0224(C)\tablefootmark{b}\\  
        HARPSpre                   & Lagrange    &6&192.C-0224(H)\tablefootmark{b}\\
        HARPSpre/post                    & Diaz&9&096.C-0499(A)\tablefootmark{b}\\
        HARPSpost                   &Lagrange&8&098.C-0739(A)\tablefootmark{b}\\
        HARPSpost                   &Lagrange&2&099.C-0205(A)\tablefootmark{b}\\
        HARPSpost                   &Diaz&11&0100.C-0487(A)\tablefootmark{b}\\
        HARPSpost                   &Diaz&4&0102.C-0525(A)\tablefootmark{b}\\
        HARPSpost                   &Bonfils&14&1102.C-0339(A)\\
        
        HARPSpre                   &Anglada-Escude&90&191.C-0505(A)\tablefootmark{b}\\
        HARPSpost                   &Berdinas&402&0101.D-0494(B)\tablefootmark{b}\\
        
        \textbf{HARPSpost}                   &\textbf{Jeffers}&\textbf{60}&\textbf{0101.C-0516(A)}\tablefootmark{1}\\
        \textbf{HARPSpost}                   &\textbf{Jeffers}&\textbf{70}&\textbf{0104.C-0863(A)}\\
        ESPRESSO                   &Pepe&7&1102.C-0744\\
        ESPRESSO                  &Pepe&12&1104.C-0350\\
        
        \textbf{HARPSpw}                   &\textbf{Jeffers}&\textbf{22}&\textbf{2107.C-5023(A)}\\
        \textbf{HARPSpw}                   &\textbf{Jeffers}&\textbf{20}&\textbf{0108.C-0235(B)}\\
        \textbf{HARPSpw}                    &\textbf{Jeffers}&\textbf{10}&\textbf{0108.C-0235(C)}\\

        \hline
    \end{tabular}
    
    \tablefoot{\tablefoottext{a}{Dataset column indicates before fibre-upgrade "pre;" after fibre-upgrade "post;" and post covid warm-up, "pw." Boldface indicates RedDots proposals.}
    \tablefoottext{b}{\cite{Jeffers2020}}}
    \label{tab:RVproposals}
\end{table}

\begin{figure}[h]
    \centering
    \includegraphics[width=0.49\textwidth]{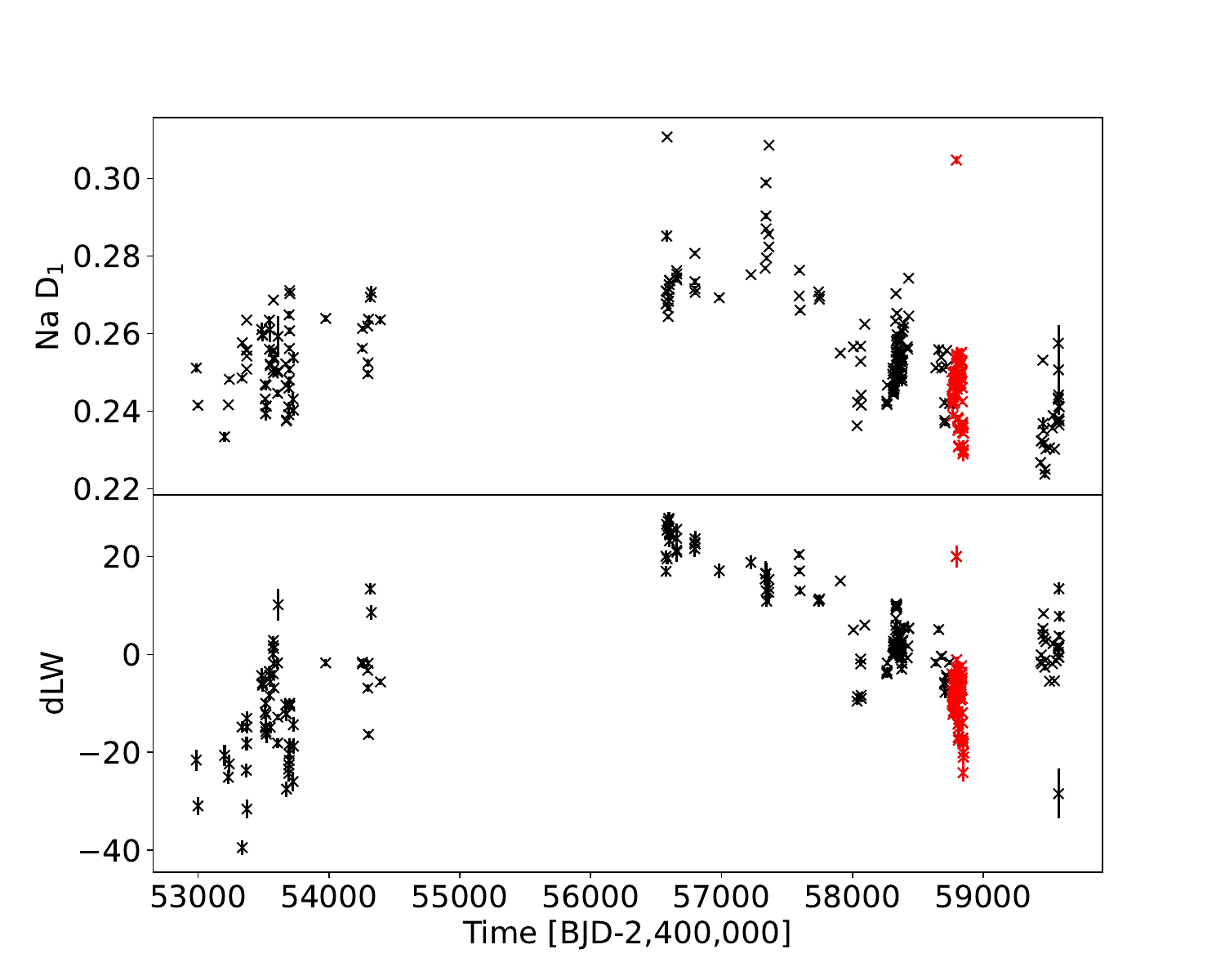}
    \caption{Long-term trend in Na D$_1$ and dLW indicators over the entire dataset. The phase where the star was active and observed over two full rotations is marked in red}
    \label{fig:longtermtrend}
\end{figure}

\begin{table}
    \centering
    \caption{Priors for all QP models with candidates among \textit{f}, \textit{e}, \textit{b}, \textit{c,} and \textit{d}\tablefootmark{a}.}
    \begin{tabular}{ll}
    \hline Parameter & Prior\\
    \hline\hline
    $P_f$ & $\mathcal{U}$(2.21,2.22)\\
    $t_{0,f}$ & $\mathcal{U}$(2453485,2453487.5)\\
    $K_f$ & $\mathcal{U}$(0.01,5)\\
    $e_f$ & $\mathcal{U}$(0,0.9)\\
    $\omega_f$ & $\mathcal{U}$(0,$2\pi$)\\
    $P_e$ & $\mathcal{U}$(4.42,4.43)\\
    $t_{0,e}$ & $\mathcal{U}$(2453486,2453491)\\
    $K_e$ & $\mathcal{U}$(0.01,5)\\
    $e_e$ & $\mathcal{U}$(0,0.9)\\
    $\omega_e$ & $\mathcal{U}$(0,$2\pi$)\\
    $P_b$ & $\mathcal{U}$(9.24,9.28)\\
    $t_{0,b}$ & $\mathcal{U}$(2453488,2453498)\\
    $K_b$ & $\mathcal{U}$(0.01,5)\\
    $e_b$ & $\mathcal{U}$(0,0.9)\\
    $\omega_b$ & $\mathcal{U}$(0,$2\pi$)\\
    $P_c$ & $\mathcal{U}$(21.6,21.9)\\
    $t_{0,c}$ & $\mathcal{U}$(2453484,2453506)\\
    $K_c$ & $\mathcal{U}$(0.01,5)\\
    $e_c$ & $\mathcal{U}$(0,0.9)\\
    $\omega_c$ & $\mathcal{U}$(0,$2\pi$)\\
    $P_d$ & $\mathcal{U}$(50.5,51)\\
    $t_{0,d}$ & $\mathcal{U}$(2453484,2453536)\\
    $K_d$ & $\mathcal{U}$(0.01,5)\\
    $e_d$ & $\mathcal{U}$(0,0.9)\\
    $\omega_d$ & $\mathcal{U}$(0,$2\pi$)\\
    $\mu_{\mathrm{pre}} \mathrm{[m/s]}$ & $\mathcal{U}$(-30,30)\\
    $\mu_{\mathrm{post}} \mathrm{[m/s]}$ & $\mathcal{U}$(-30,30)\\
    $\mu_{\mathrm{pw}} \mathrm{[m/s]}$ & $\mathcal{U}$(-30,30)\\
    $\sigma_{w,\mathrm{pre}} \mathrm{[m/s]}$ & $\mathcal{J}$(0.01,10)\\
    $\sigma_{w,\mathrm{post}} \mathrm{[m/s]}$ & $\mathcal{J}$(0.01,10)\\
    $\sigma_{w,\mathrm{pw}} \mathrm{[m/s]}$ & $\mathcal{J}$(0.01,10)\\
    $\sigma_{\mathrm{GP,pre}} \mathrm{[m/s]}$ & $\mathcal{J}$(0.01,10.0)\\
    $\sigma_{\mathrm{GP,post}} \mathrm{[m/s]}$ & $\mathcal{J}$(0.01,10.0)\\
    $\sigma_{\mathrm{GP,pw}} \mathrm{[m/s]}$ & $\mathcal{J}$(0.01,10.0)\\
    $P_{\mathrm{rot,GP}} [days]$ & $\mathcal{J}$(10,100)\\
    $\Gamma_{\mathrm{GP}}$ & $\mathcal{J}$(0.1,1000)\\
    $\alpha_{\mathrm{GP}} [\mathrm{days}^{-2}]$ & $\mathcal{J}$($10^{-8}$,1)\\
    \hline
    \end{tabular}
    \tablefoot{\tablefoottext{a}{$\mathcal{U}$ indicates a uniform prior and $\mathcal{J}$ indicates a log-uniform prior. When a circular orbit was fitted $e=0$ and $\omega=\pi/2$ was fixed for the specific planet.}}
    \label{tab:priors}
\end{table}

\begin{table}[]
    \centering
    \caption{First RVs of the HARPS dataset after binning.}
    \begin{tabular}{lll}
    \hline
    \textbf{BJD [d]} & \textbf{RV [$\mathrm{m\,s^{-1}}$]} & \boldmath$\sigma_{\mathrm{RV}}$ \textbf{[$\mathrm{m\,s^{-1}}$]}\\
    \hline\hline
    2452985.571688 & 3.119 & 0.819  \\ 
    2452998.538683 & 2.643 & 0.484  \\ 
    2453201.899879 & 1.230 & 0.790\\ 
    2453230.766890 & 10.238 & 0.330  \\ 
    2453237.873496 & 2.387 & 0.365  \\
    ...&...&...\\
\hline
    \end{tabular}

    \label{tab:RVdata}
\end{table}

\begin{figure}
    \centering
    \includegraphics[width=0.49\textwidth]{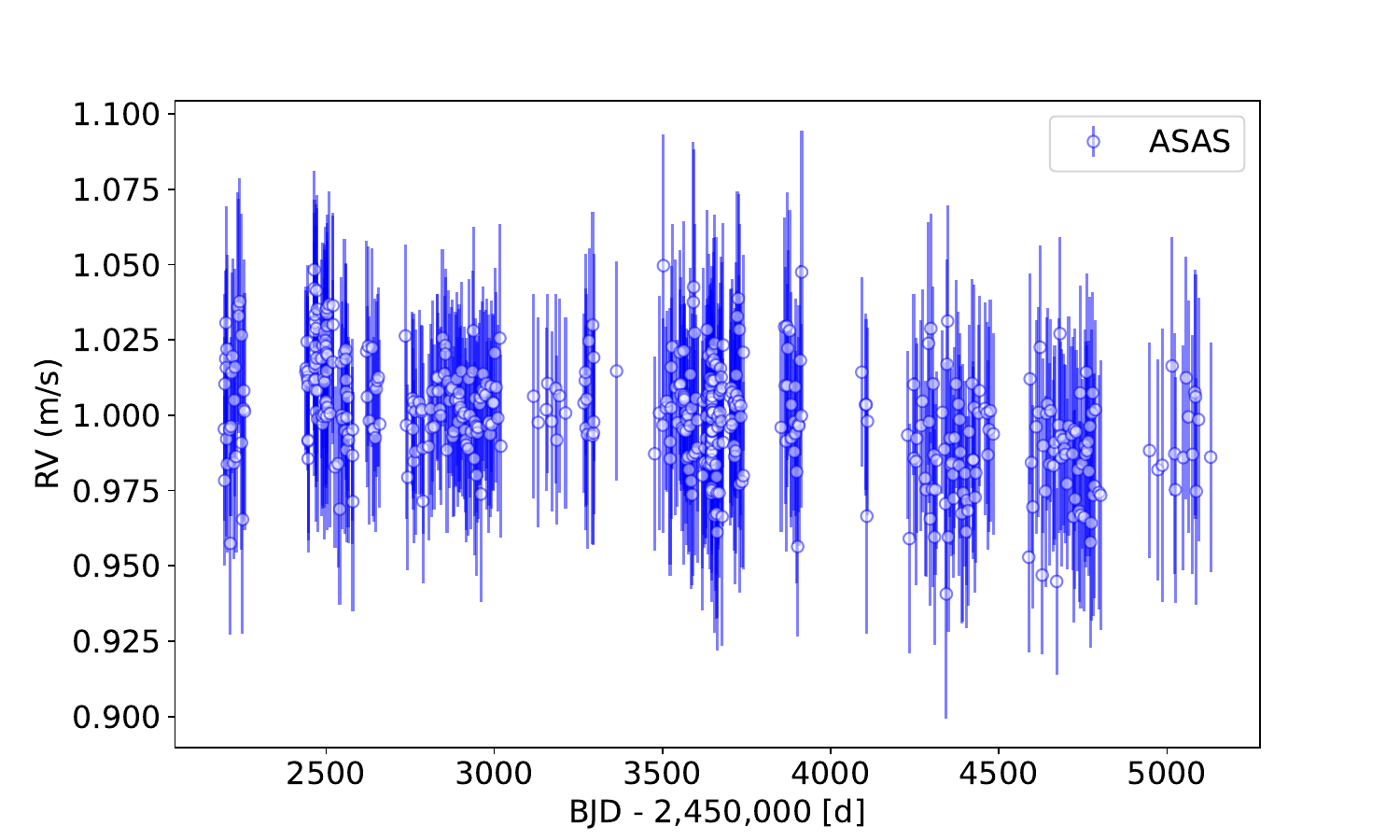}
    \caption{Photometric time series of GJ 887 from the ASAS survey}
    \label{fig:ASASTS}
\end{figure}

\begin{table}[h]
    \centering
    \caption{Same as Table \ref{tab:priors} with the priors for period and $t_0$ for injected planets\tablefootmark{a}.}
    \begin{tabular}{ll}
    \hline Parameter & Prior\\
    \hline\hline
    $P_{3}$ & $\mathcal{U}$(3.095,3.105)\\
    $t0_{3}$ & $\mathcal{U}$(2453485,2453488.2)\\
    $P_{66}$ & $\mathcal{U}$(65.7,66.3)\\
    $t0_{66}$ & $\mathcal{U}$(2453484,2453561)\\

    \hline
    \end{tabular}
    \tablefoot{\tablefoottext{a}{The other priors for these planets were $\pi(K)=\mathcal{U}$(0.01,5), $\pi(e)=\mathcal{U}$(0,0.9) and $\pi(\omega)=\mathcal{U}(0,2\pi)$.}}

    \label{tab:priorsinj}
\end{table}

\begin{figure}[h]
    \centering
    \includegraphics[width=0.49\textwidth]{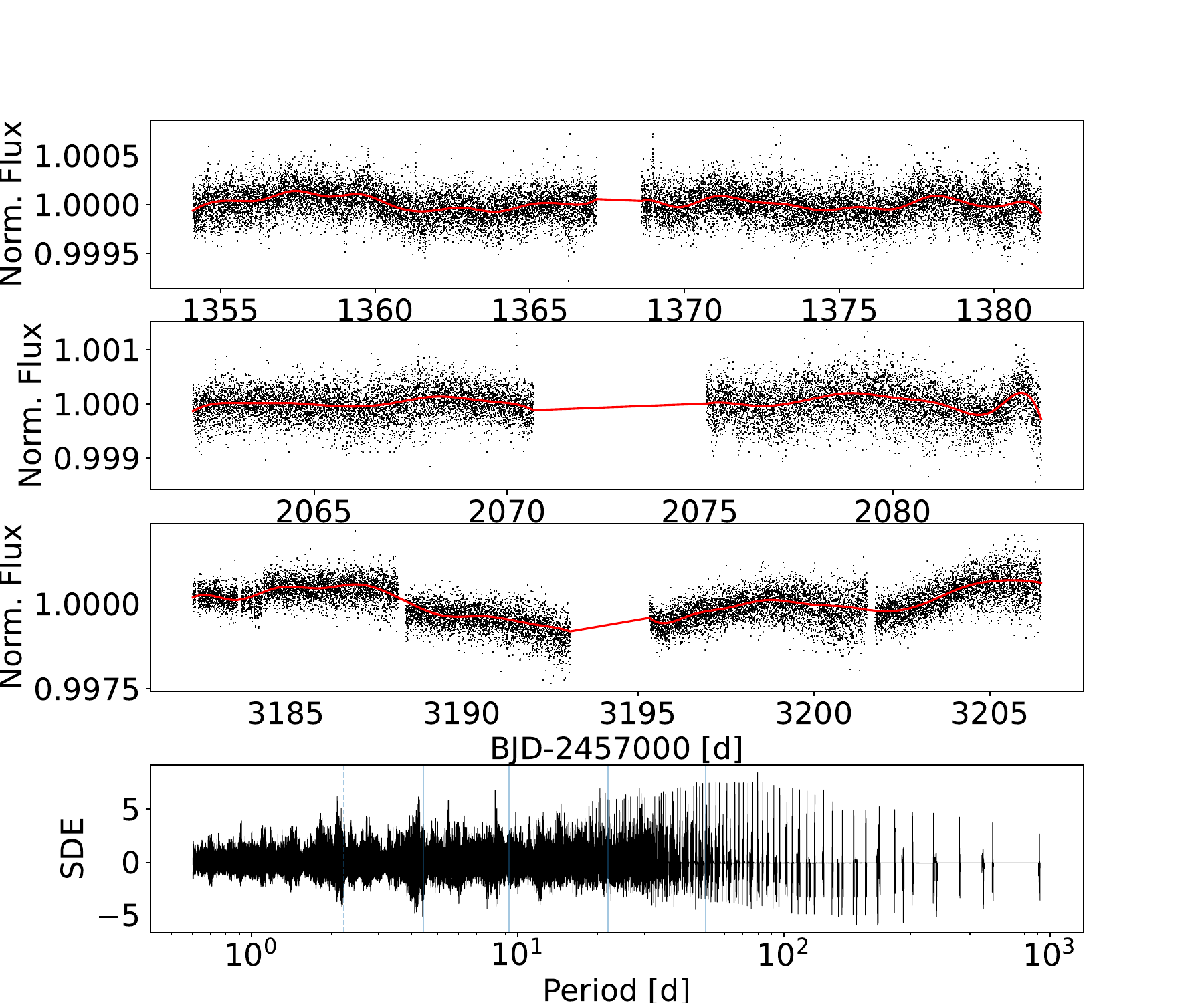}
    \caption{TESS PDCSAP light curves of GJ 887 of sectors 2, 28, and 69 in the top three panels with trends fitted as splines in red. In the lower panel a Transit Least Squares periodogram is shown along with vertical bars for the location of the planets (\textit{solid}) and the candidate (\textit{dashed}).}
    \label{fig:Tsearch}
\end{figure}

\begin{figure*}
    \centering
    \includegraphics[width=\textwidth]{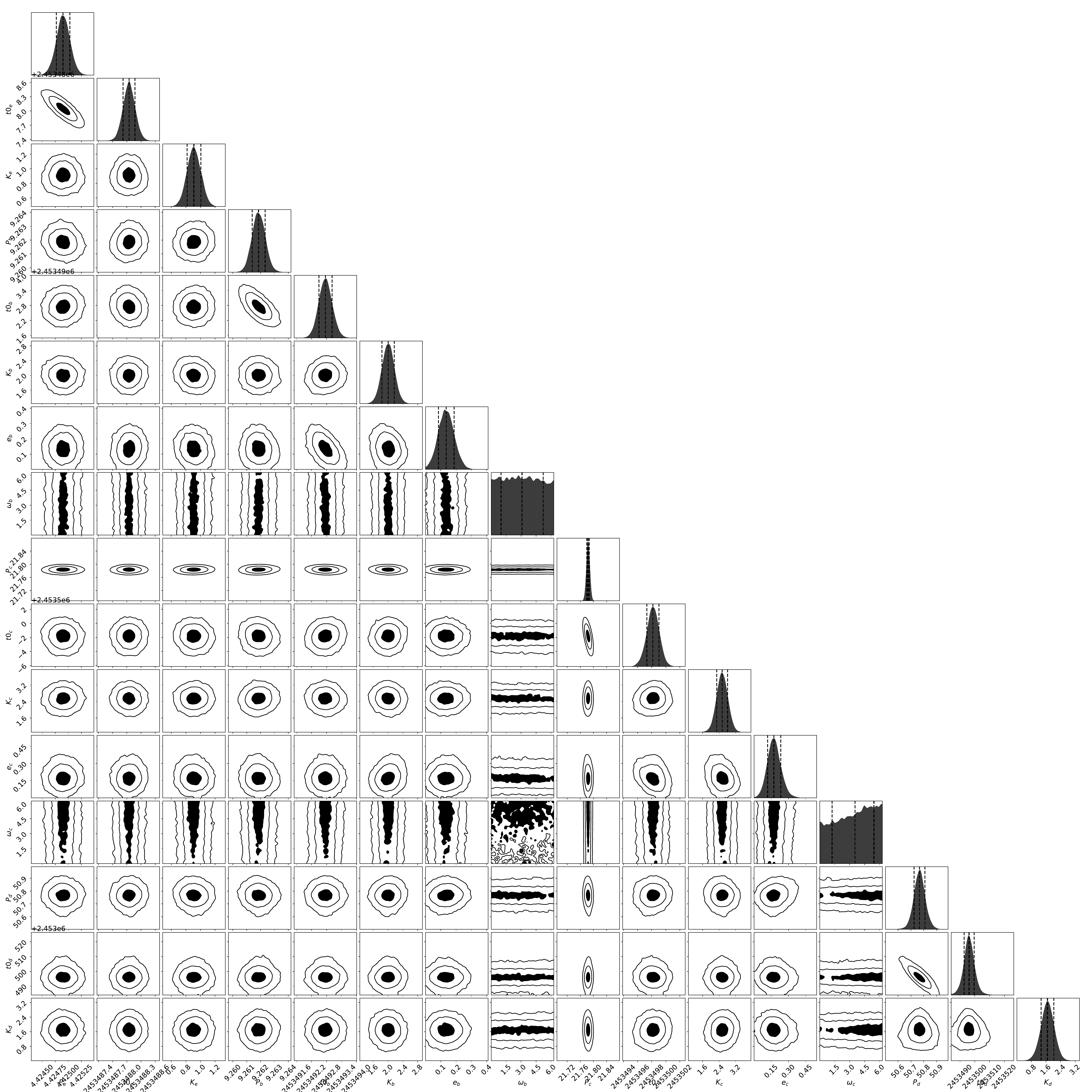}
    \caption{Correlations between orbital parameters of the planets in the four-planet model including planets e, b, c, and d.}
    \label{fig:cornerplanet}
\end{figure*}

\begin{figure*}
    \centering
    \includegraphics[width=0.6\textwidth]{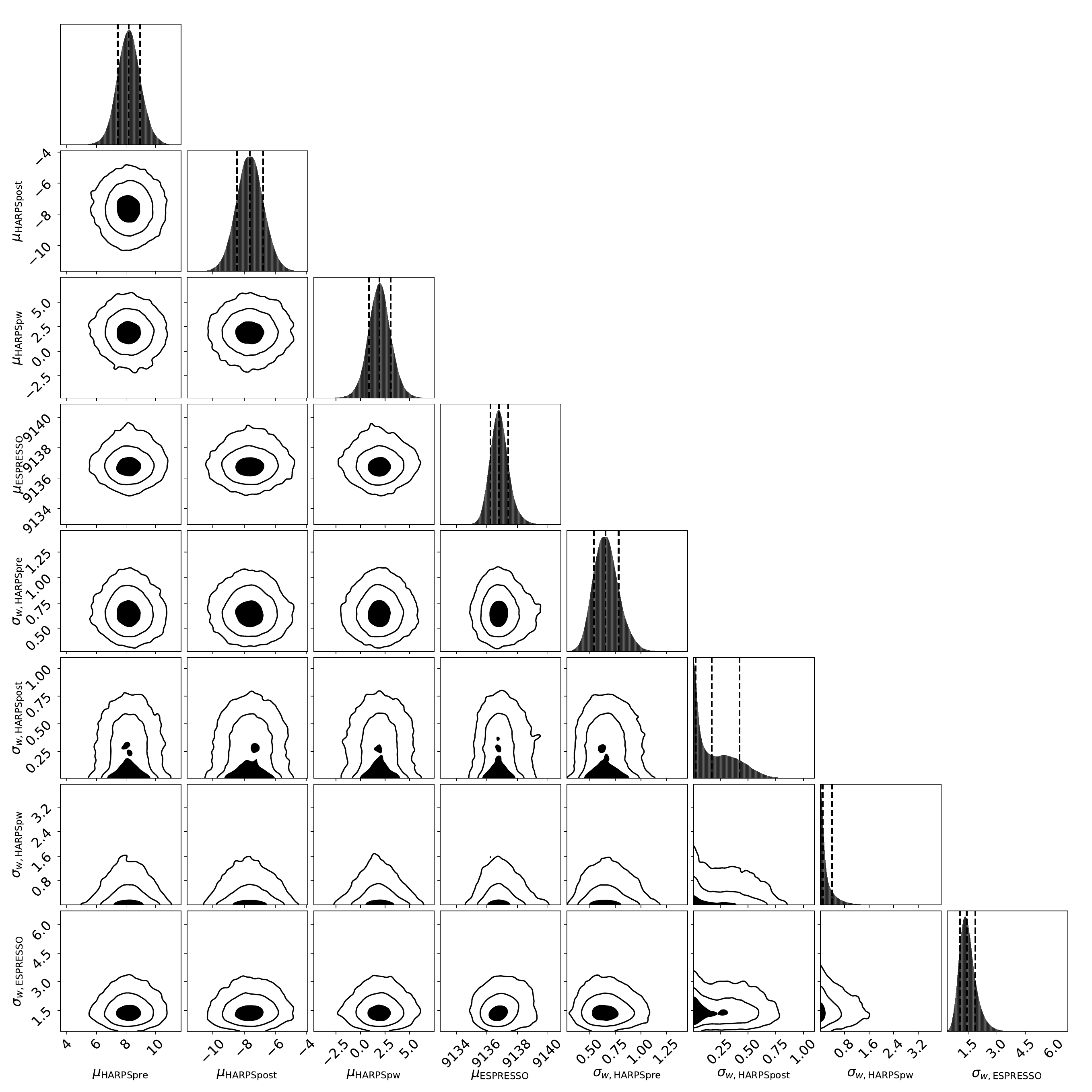}
    \caption{Same as Fig. \ref{fig:cornerplanet}, but for jitters and offsets}
    \label{fig:cornerRV}
\end{figure*}

\begin{figure*}
    \centering
    \includegraphics[width=0.6\textwidth]{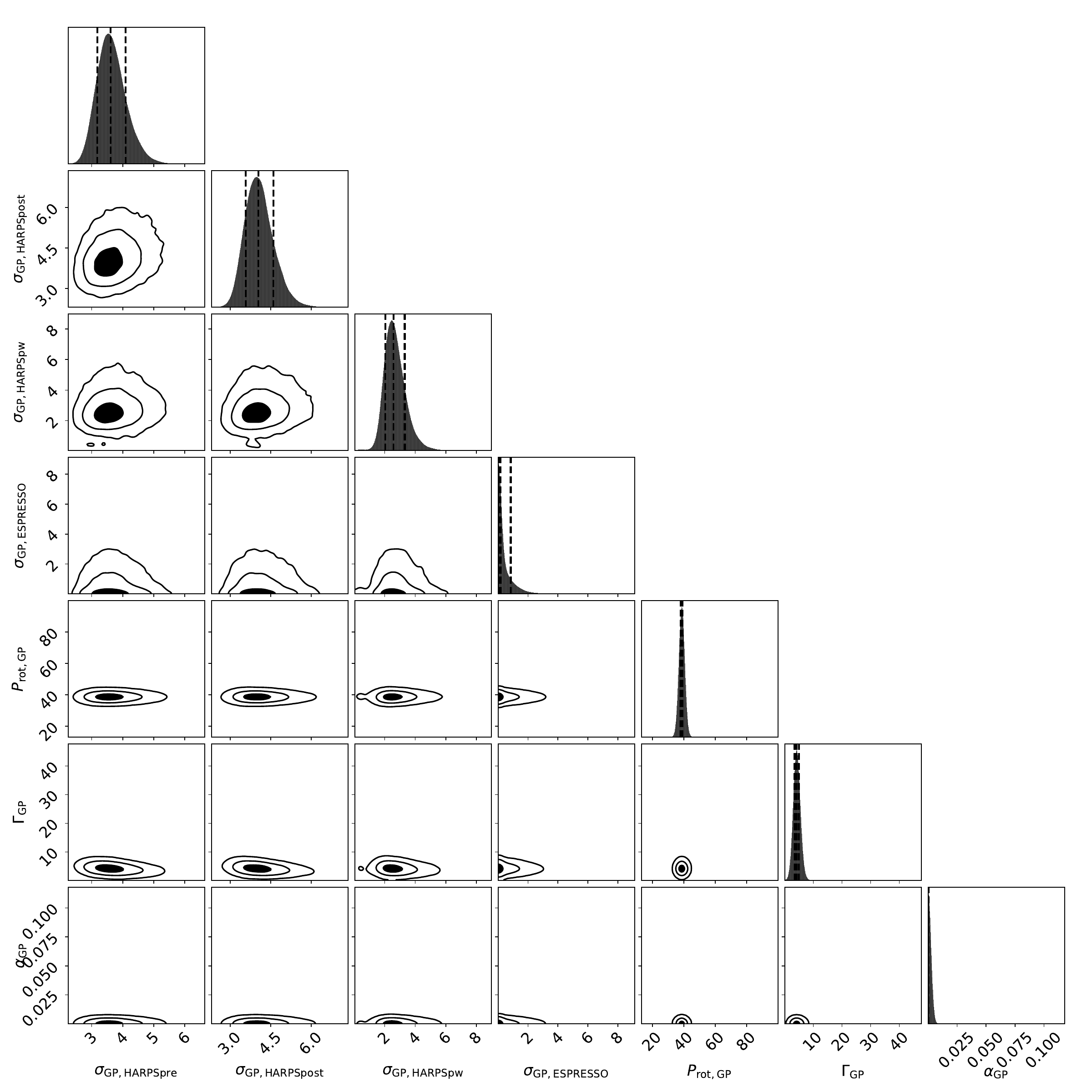}
    \caption{Same as Fig. \ref{fig:cornerplanet}, but for the GP parameters.}
    \label{fig:cornerGP}
\end{figure*}

\begin{table*}[]
    \centering
    \caption{$\log\mathcal{Z}$ values of all GP models\tablefootmark{a}.}
    \begin{tabular}{lllll}
        \hline Model &  $\log\mathcal{Z}$ QP &  $\Delta\log\mathcal{Z}$ QP &  $\log\mathcal{Z}$ dSHO &  $\Delta\log\mathcal{Z}$ dSHO    \\
            \hline
            \hline  
    \textit{2c,4c,9e,21e,51c} (E) &   -$640.05\pm1.04$ &        +4.42 &     -645.06 &         -0.59  \\
    \textbf{\textit{4c,9e,21e,51c} (D)} &   \textbf{-$644.47\pm0.84$} &        \textbf{0.00} &     \textbf{-645.8} &         \textbf{-1.33}  \\
    \textit{2e,4c,9e,21e,51c} &   -644.82 &        -0.35 &     -646.91 &        -2.44  \\
    \textit{4e,9e,21e,51c} &   -645.98 &        -1.53 &     -649 &        -4.53 \\
    
    \textit{2c,4c,9e,21e} (F) &   -$649.47\pm2.64$ &        -5.00 &     -648.37 &         -3.9  \\
    \textit{4c,9e,21e} (C) &   -$650.95\pm0.38$ &        -6.48 &     -651.85 &         -7.38  \\
    \textit{4e,9e,21e} &   -657.38 &        -12.91 &     -652.75 &        -8.28  \\
    \textit{9e,21e,51c} (B) &   -671.62 &        -27.15 &     -669.21 &         -24.74  \\
    \textit{9e,21e,51e} &   -676.29 &        -31.82 &     -669.89 &         -25.42  \\
    \textit{9e,21e} (A) &   -678.87 &        -34.4 &     -677.51 &        -33.04  \\
    \textit{9e,21c} &   -681.77 &        -37.3 &     -681.32 &         -36.85  \\
    \textit{9e} &   -692.95 &        -48.48 &     -688.48 &         -44.01  \\
    \textit{9c} &   -695.72 &        -51.25 &     -692.94 &        -48.47  \\
    \textit{0P}  &   -727.51 &        -83.04 &     -723.52 &         -79.05  \\    
    \hline
    \end{tabular}
    \tablefoot{\tablefoottext{a}{$\Delta\log z$ calculated in respect to model D with QP kernel. The letter \textit{c} indicates a circular orbit, \textit{e} an elliptical orbit.}}
    \label{tab:results}
\end{table*}

\begin{table*}
\centering
    \caption{Same as Table \ref{tab:results4p} for five-planet model.}
\label{tab:results5p}

\begin{tabular}{llllll}
\hline Parameter & candidate \textit{f} &planet \textit{e} &planet \textit{b} &planet \textit{c} &planet \textit{d} \\
\hline\hline

$P$ [days] & $2.21661_{-0.00009}^{+0.00010}$ & $4.42494_{-0.00013}^{+0.00012}$ & $9.2619_{-0.0005}^{+0.0005}$ & $21.784_{-0.004}^{+0.004}$ & $50.77_{-0.04}^{+0.05}$ \\
$K \mathrm{[m/s]}$ & $0.37_{-0.09}^{+0.08}$ & $0.92_{-0.09}^{+0.09}$ & $2.05_{-0.16}^{+0.16}$ & $2.54_{-0.27}^{+0.27}$ & $1.7_{-0.4}^{+0.4}$ \\
$t_0$ [BJD] & $2453486.46_{-0.19}^{+0.18}$ & $2453488.04_{-0.12}^{+0.13}$ & $2453492.73_{-0.25}^{+0.25}$ & $2453498.2_{-0.9}^{+0.9}$ & $2453496_{-4}^{+4}$ \\
$e$ & - & - & $0.13_{-0.05}^{+0.05}$ & $0.17_{-0.06}^{+0.06}$ & - \\
$\omega$ [rad] & - & - & $2.9_{-2.0}^{+2.3}$ & $3.4_{-2.3}^{+2.0}$ & - \\
$M \sin i\,\, [M_{\oplus}]$ & $0.47_{-0.11}^{+0.11}$ & $1.48_{-0.17}^{+0.18}$ & $4.1_{-0.5}^{+0.5}$ & $6.5_{-0.9}^{+1.0}$ & $6.1_{-1.3}^{+1.4}$ \\
$a$ [AU] & $0.0263_{-0.0009}^{+0.0009}$ & $0.0417_{-0.0015}^{+0.0014}$ & $0.0683_{-0.0024}^{+0.0022}$ & $0.121_{-0.005}^{+0.004}$ & $0.212_{-0.008}^{+0.007}$ \\
$T_{\mathrm{eq}}$ [K] & $685_{-25}^{+26}$ & $544_{-20}^{+21}$ & $426_{-16}^{+16}$ & $320_{-12}^{+12}$ & $241_{-9}^{+9}$ \\
$S$ [$S_{\oplus}$] & $53_{-8}^{+9}$ & $20.9_{-2.9}^{+4.0}$ & $7.8_{-1.1}^{+1.3}$ & $2.5_{-0.4}^{+0.4}$ & $0.81_{-0.12}^{+0.13}$ \\
$\theta$ [mas] & $8.01_{-0.28}^{+0.26}$ & $12.7_{-0.5}^{+0.5}$ & $20.8_{-0.8}^{+0.7}$ & $36.7_{-1.3}^{+1.2}$ & $64.6_{-2.2}^{+2.1}$ \\
$C \times 10^{-9}$ & $258_{-16}^{+19}$ & $200_{-13}^{+15}$ & $198_{-13}^{+15}$ & $118_{-8}^{+9}$ & $32.8_{-2.1}^{+2.4}$ \\
\hline
    \multicolumn{6}{c}{\textit{RV Parameters}} \\
    \hline
\textit{$\mu_{\mathrm{HARPSpre}} \mathrm{[m/s]}$} & $8.1_{-0.8}^{+0.8}$ & - & - & - & - \\
\textit{$\sigma_{w,\mathrm{HARPSpre}} \mathrm{[m/s]}$} & $0.56_{-0.12}^{+0.14}$ & - & - & - & - \\
\textit{$\mu_{\mathrm{HARPSpost}} \mathrm{[m/s]}$} & $-7.7_{-0.9}^{+0.9}$ & - & - & - & - \\
\textit{$\sigma_{w,\mathrm{HARPSpost}} \mathrm{[m/s]}$} & $0.08_{-0.06}^{+0.20}$ & - & - & - & - \\
\textit{$\mu_{\mathrm{HARPSpw}} \mathrm{[m/s]}$} & $1.8_{-1.2}^{+1.1}$ & - & - & - & - \\
\textit{$\sigma_{w,\mathrm{HARPSpw}} \mathrm{[m/s]}$} & $0.08_{-0.06}^{+0.22}$ & - & - & - & - \\
\textit{$\mu_{\mathrm{ESPRESSO}} \mathrm{[m/s]}$} & $9136.8_{-0.6}^{+0.7}$ & - & - & - & - \\
\textit{$\sigma_{w,\mathrm{ESPRESSO}} \mathrm{[m/s]}$} & $1.6_{-0.4}^{+0.5}$ & - & - & - & - \\
\hline
    \multicolumn{6}{c}{\textit{GP Parameters}} \\
    \hline
\textit{$\sigma_{\mathrm{GP,HARPSpre}} \mathrm{[m/s]}$} & $3.8_{-0.5}^{+0.6}$ & - & - & - & - \\
\textit{$\sigma_{\mathrm{GP,HARPSpost}} \mathrm{[m/s]}$} & $4.1_{-0.5}^{+0.6}$ & - & - & - & - \\
\textit{$\sigma_{\mathrm{GP,HARPSpw}} \mathrm{[m/s]}$} & $2.5_{-0.6}^{+0.7}$ & - & - & - & - \\
\textit{$\sigma_{\mathrm{GP,ESPRESSO}} \mathrm{[m/s]}$} & $0.14_{-0.12}^{+0.60}$ & - & - & - & - \\
\textit{$P_{\mathrm{rot,GP}} [\mathrm{days}^{-2}]$} & $38.7_{-0.5}^{+0.6}$ & - & - & - & - \\
\textit{$\Gamma_{\mathrm{GP}}$} & $4.1_{-0.7}^{+0.8}$ & - & - & - & - \\
\textit{$\alpha_{\mathrm{GP}} [\mathrm{days}^{-2}]$} & $0.00029_{-0.00009}^{+0.00015}$ & - & - & - & - \\

\hline\end{tabular}
\end{table*}

\begin{figure*}[h]
    \centering
    \includegraphics[width=0.9\textwidth]{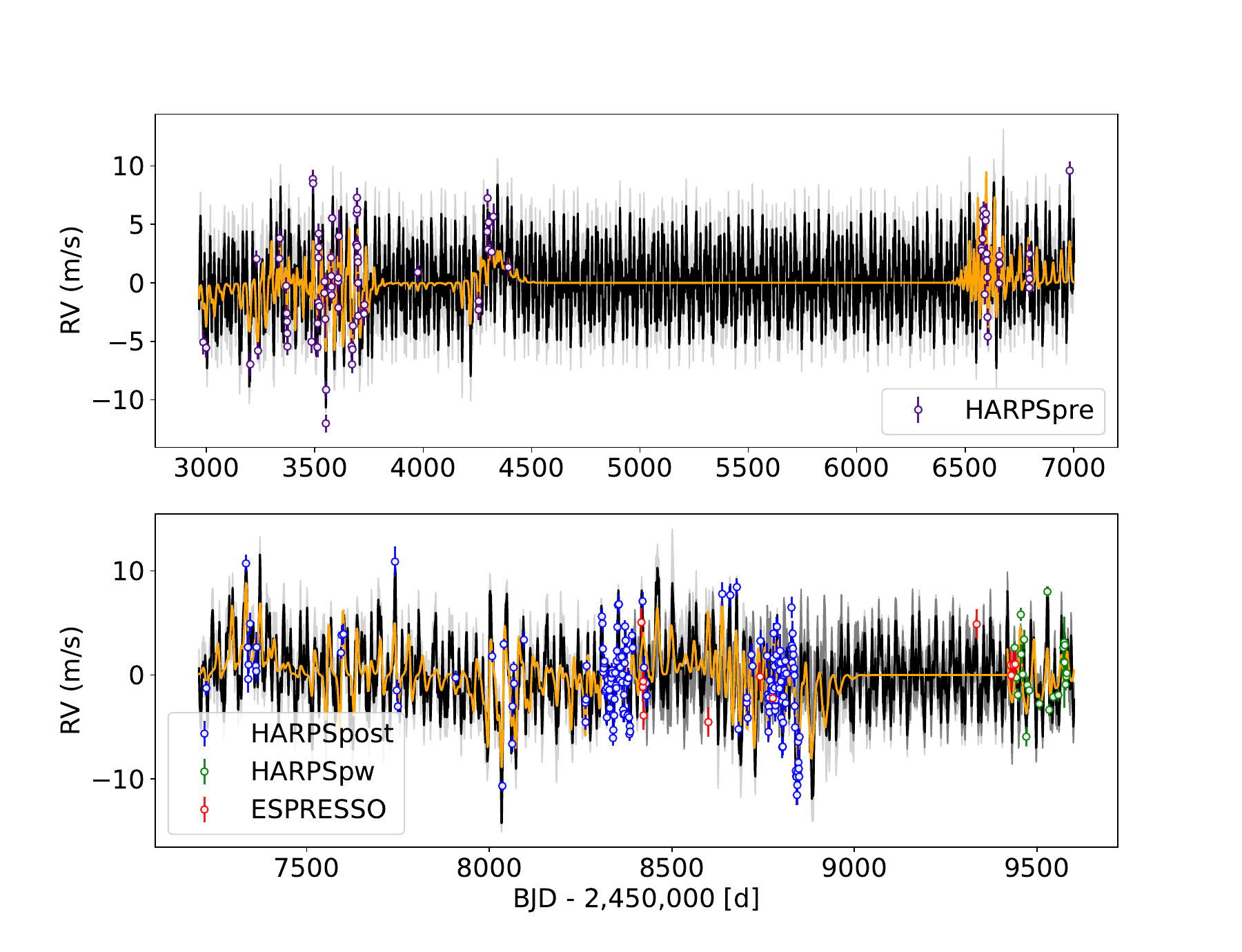}
    \caption{Four-planet RV model of planets e, b, c, and d including Quasi-periodic GP (\textit{black}) with 95\%-confidence interval (\textit{grey}) and GP contribution (\textit{yellow})}
    \label{fig:RVmodel}
\end{figure*}
\begin{figure*}
    \centering
    \includegraphics[width=0.9\textwidth]{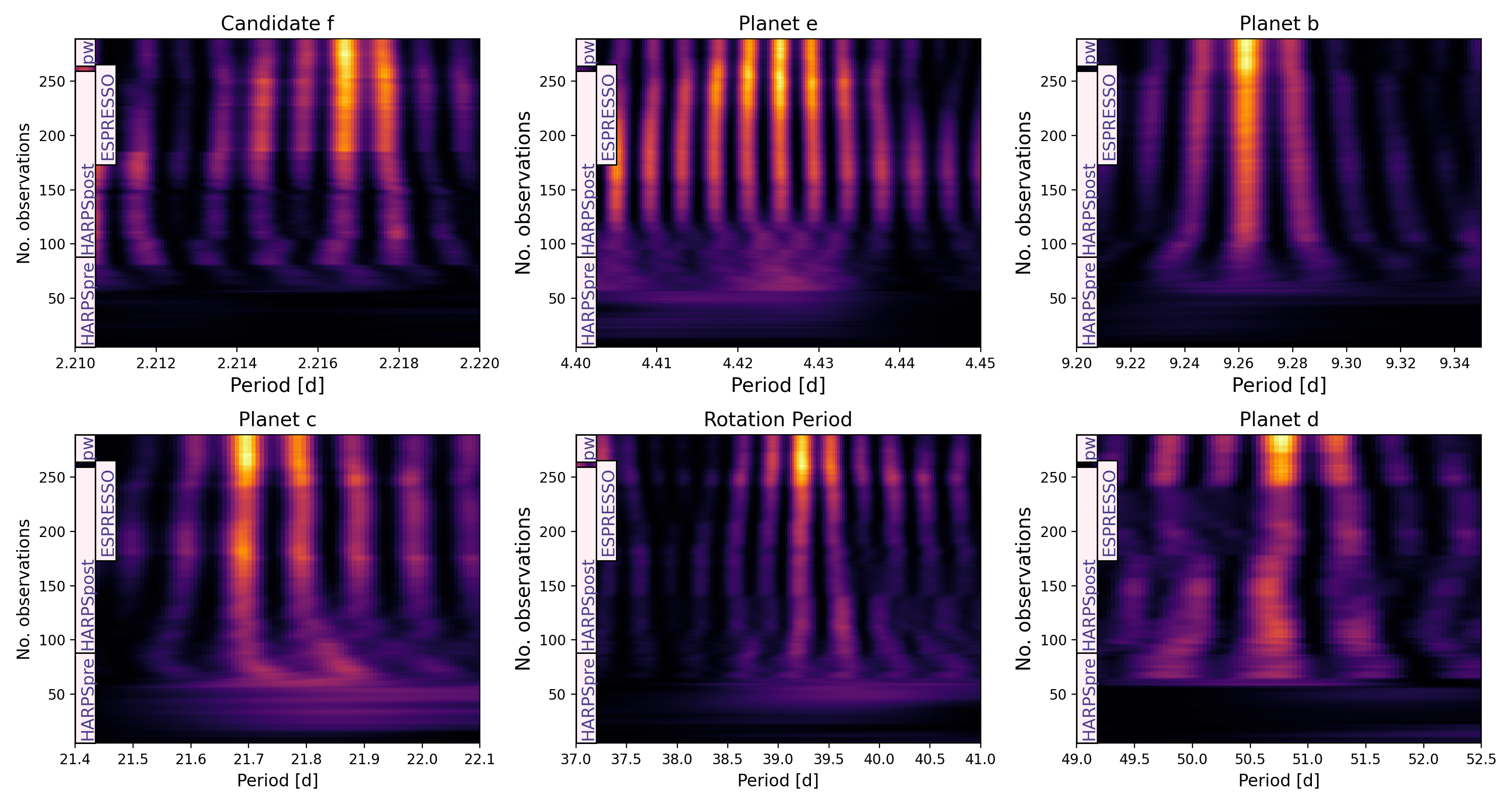}
    \caption{sBGLS periodograms of all planet candidates and the rotation period of the star. The apparent fringe pattern in all panels is caused by the sampling of the data in two chunks separated by approximately $13~\mathrm{y}$.}
    \label{fig:sbgls}
\end{figure*}

\end{appendix}

\end{document}